\documentclass[draftcls,onecolumn]{IEEEtran}

\usepackage{cite}
\usepackage[cmex10]{amsmath}
\usepackage{amssymb}
\usepackage{array}
\usepackage{graphicx}
\usepackage[caption=false,font=footnotesize]{subfig}
\usepackage{color}

\newtheorem{theorem}{Theorem}
\newtheorem{lemma}{Lemma}
\newtheorem{remark}{Remark}

\newtheorem{proposition}{Proposition}
\newcommand\relphantom[1]{\mathrel{\phantom{#1}}}

\begin{document}
\title{Distributed Inference for Relay-Assisted Sensor Networks With Intermittent Measurements Over Fading Channels\thanks{This research is funded by the Republic of Singapore's National Research Foundation through a grant to the Berkeley Education Alliance for Research in Singapore (BEARS) for the Singapore-Berkeley Building Efficiency and Sustainability in the Tropics (SinBerBEST) Program. BEARS has been established by the University of California, Berkeley as a center for intellectual excellence in research and education in Singapore. }
}

\author{Shanying~Zhu, Yeng Chai Soh,~\IEEEmembership{Senior~Member,~IEEE,} and~Lihua Xie,~\IEEEmembership{Fellow,~IEEE} \\
\thanks{
S. Zhu is with EXQUISITUS,  School of Electrical and Electronic Engineering, Nanyang Technological University, 50 Nanyang Avenue, Singapore 639798, and also with Department of Automation, Shanghai Jiao Tong University,
Shanghai 200240, China, and the Key Laboratory of System Control and
Information Processing, Ministry of Education of China, Shanghai 200240,
China
({\tt e-mail:shyingzhu@gmail.com})}
\thanks{Y. C. Soh and L. Xie are with EXQUISITUS,  School of Electrical and Electronic Engineering, Nanyang Technological University, 50 Nanyang Avenue, Singapore 639798
({\tt e-mail:  eycsoh@ntu.edu.sg; elhxie@ntu.edu.sg}).}
}

 \markboth{A short version is to appear in IEEE Transactions on Signal Processing}%
 {S. Zhu, Y. C. Soh,  L. Xie: Distributed estimation in sensor networks}
%

\maketitle

\begin{abstract}
  In this paper, we consider a general distributed estimation  problem  in relay-assisted sensor networks by taking into account time-varying asymmetric communications, fading channels and intermittent measurements. Motivated by centralized filtering algorithms, we propose a distributed innovation-based estimation algorithm by combining the measurement innovation (assimilation
  of new measurement) and local data innovation (incorporation of neighboring data). Our algorithm is fully distributed which does not need a fusion center. We establish theoretical results regarding asymptotic unbiasedness and consistency of the proposed algorithm. Specifically, in order to cope with time-varying asymmetric communications, we utilize an ordering technique and the generalized Perron complement to manipulate the first and second moment analyses in a tractable framework. Furthermore, we present a performance-oriented design of the proposed algorithm for energy-constrained networks based on the theoretical results. Simulation results corroborate the theoretical findings, thus demonstrating the effectiveness of the proposed algorithm.
\end{abstract}


\IEEEpeerreviewmaketitle

\section{Introduction}
 A sensor network is composed of a large number of  nodes in which each node has limited capabilities of sensing, data processing and communication. But as a network, it is able to perform some desired sensing tasks, among which estimation is one basic application.
 In a typical estimation problem over sensor networks, nodes make noisy measurements of a scalar parameter of interest.  The main concern is how to utilize the measurements to produce a desired estimate by only exchanging information between neighboring  nodes. 

 One important aspect of sensor networks that should be considered is that nodes usually have limited energy and computational abilities.  Recently, there have been considerable interests in distributed estimation schemes for sensor networks to utilize these resources efficiently \cite{CatSay10,DasMes09,KarMouRam12,ScuBarPes08,SchRibGia08,SchRibGia082,StanSti11,XiaoBoyLal05,XiaRibLuo_etal06}. Compression and quantization before data transmission at the node level is one way to reduce the amount of transmission. And several forms of quantizers have been used for distributed parameter estimation problems \cite{FangLi10,LiFuXieZhang11,LiuLiXieetal13,KarMouRam12}. 
 Although quantization errors are inevitably incurred, convergence can still be guaranteed in the presence of symmetric or balanced communications.

  For large-scale field monitoring, the nodes would often be separated into several groups after initial deployment. In this scenario, compression and quantization at the node level would no longer be an energy efficient way, since long-distance communication is more energy-consuming \cite{KarlWill05}. This problem has motivated the use of static/mobile relays  to connect these separate groups \cite{DiFranDasAna11,HanCaoetal10,WangXuetal07},
  such that the communications between nodes are confined to shorter distances, while meeting certain network specifications, 
   which leads to the notion of relay-assisted sensor network. 
  The benefit is that with only a few relays, the degree of network connectivity  can be increased (see Fig.~\ref{fig:example} (left)). Moreover, long-distance communications can be avoided, thus achieving better energy efficiency.

  Although much research has been done on relay-assisted sensor networks with regard to node deployment, improvement of energy balance and so on, few results have been reported for solving distributed estimation problems in such networks. In this work, we focus on the estimation problem for such relay-assisted sensor networks with sensor nodes (SNs) and relay nodes (RNs). In our setting, SNs make measurements and do some fusion operations after collecting data from neighboring nodes; they are the endpoints of messages. Whereas RNs are not the endpoints of communication, they only act as forwarders, i.e., RNs carry the collected data, until they get in contact with SNs. Ref. \cite{DiFranDasAna11} presents a recent survey for scenarios using mobile RNs to collect data. Rather than the centralized approaches in \cite{DiFranDasAna11}, we are more interested in the distributed schemes, where there are no sinks or fusion centers. All SNs cooperate with their neighbors to estimate the unknown parameter based only on local transmission.

  During the information  exchange, the nodes use the analog-and-forward scheme, where data are scaled and then transmitted without any coding  \cite{LiuGamaSaye07,LeongDeyEvans11}. This scheme is attractive due to its simplicity as well as the possibility of real-time  processing since there is no coding delay. A number of distributed estimation algorithms based on this analog forwarding have been proposed for sensor networks without RNs. For instance, the authors of \cite{DasMes09,OlfFaxMur07,PerZam10,ScuBarPes08,SchRibGia08,XiaoBoyLal05} proposed  use of distributed consensus algorithms for sensor fusion, where the nodes first take measurements and then start consensus algorithms to fuse the data to achieve the average or weighted average of all the  measurements. 
  These algorithms are appropriate for the scenarios where the measurement rate is much slower than the communication rate between nodes.
  An alternative is to combine the consensus step and innovation step together if streaming measurements  can be obtained at the nodes\cite{KarMou11,KarMouRam12,ZhangZhang12}. Both methods can be shown to guarantee some kinds of convergence under certain conditions, e.g., asymptotic convergence for noise-free case, mean-square and almost sure convergence in the presence of measurement and communication noises. 
  We note that most works assumed that the communication topology is symmetric or balanced. 
  This assumption significantly simplifies the mathematical analysis. Moreover, 
  However, the symmetric or balanced requirement is restrictive and imposes much computational difficulties to construct the associated weight matrices, especially when the network varies with time, which is the case considered in this paper.

  In \cite{ZhuChenMaYangGuan15}, we removed the aforementioned restrictive requirements and investigated the situation monitoring in relay-assisted networks. A distributed algorithm called DCUE algorithm is proposed with certain performance guarantees. It is noted, however,  that the results in \cite{ZhuChenMaYangGuan15} only apply to time-invariant topologies with perfect channels, i.e., the nodes can receive data from their neighbors without any distortion or noise. Realistic networks suffer from noise and errors, which cause links to fail at random times \cite{KarMoura08}, resulting in time-varying topologies. Moreover, wireless transmissions are also subject to channel fading \cite{ChanSwamZhaoScag10,LiuGamaSaye07,MostMalm10}. Both the algorithm and the theoretical approach in \cite{ZhuChenMaYangGuan15} will fail in such situations.
  
  In this paper, we extend our previous work and consider a general framework of distributed estimation over rapidly-changing relay-assisted networks with occasional sensor failures, intermittent noisy measurements and fading channels. 
  Inspired by the centralized least-squares estimation and the distributed schemes in \cite{KarMou11,KarMouRam12,StanSti11,ZhangZhang12,ZhuChenMaYangGuan15}, we propose a distributed innovation-based estimation algorithm by combining the measurement innovation step and  local data innovation step together.  We emphasize that although several results of distributed consensus algorithms have been proposed for time-varying networks \cite{HuangDeyNairMan10,Huang12,OlfFaxMur07,ShiJoh13}, they cannot be used for our purpose. The argument is that their convergence  highly depends on the convexity property of consensus algorithms, which, however, does not hold for our estimation algorithm. To suppress the propagation of measurement and communication noises, we introduce a decaying weight in the algorithm, which tends to zero as time goes to infinity. This coupled with time-varying asymmetric communications is another big challenge, since it seems that a uniform positive lower bound on the interaction strength between nodes is necessary in consensus algorithms \cite{Berg81,HendTsit13,NedOlsOzdTsi09,PerZam10,TsitBertAtha86}, not to mention consensus-based estimation algorithms. To address this issue, we utilize a different technique that builds upon a reordered state space. This ordered structure allows us to investigate the convergence in a tractable framework, with which we can establish asymptotic unbiasedness and consistency of the proposed algorithm.

  Additionally, we consider the parameter design issue of the proposed algorithm to guarantee desired properties for energy-constrained sensor networks. In the case of available channel statistics at the receivers, we first give a simple form of decaying weights that serves the purpose.  Secondly, we present the design of amplification factors such that (i) the fading only contributes to an unbiased perturbation on the algorithm, and (ii) the transmit and received powers of each node are bounded irrespective of the number of nodes. In this way, we provide a  tractable framework for distributed estimation problems in energy-constrained relay-assisted sensor networks over time-varying asymmetric communications with  fading channels.

  The paper is organized as follows. In Section \ref{sec:model}, the network models under consideration and some preliminaries  are given. In Section \ref{sec:algorithm}, we describe a distributed innovation-based estimation algorithm. In Section \ref{sec:convergence}, we provide conditions under which asymptotic unbiasedness and consistency of  the proposed algorithm can be achieved. Section~\ref{sec:algorithmdesign} is devoted to the parameter design issue of the proposed algorithm with guaranteed performance. In Section \ref{sec:simulation}, simulation results are presented to demonstrate the effectiveness of the proposed algorithm, followed by the conclusion in Section~\ref{sec:conclusion}.

  \emph{Notation:} We use bold uppercase and lowercase letters to denote  matrices  and column vectors, respectively. $\mathbb{R}^{m\times n}$ denotes the set of all $m\times n$ real-valued matrices with the spectral norm $\|\cdot\|$ and the maximum row sum norm $\|\cdot\|_{\infty}$. In particular, $\mathbf{I}$ denotes the identity matrix, $\mathbf{1}$ and $\mathbf{0}$ stand for the all-one and all-zero vectors, respectively. For a matrix $\mathbf{A}$, $r_i(\mathbf{A})$ represents the $i$-th row sum, $r_{\max}(\mathbf{A})=\max_i r_i(\mathbf{A}) $ and $\rho(\mathbf{A})$ denotes its spectral radius. For a vector $\mathbf{x}$, we use $\|\mathbf{x}\|_p$ for the $l_p$-norm with $p=1,2,\infty$.

\section{Models and preliminaries}\label{sec:model}
\begin{figure}[!t]
\centering
 \includegraphics[width=8.5cm]{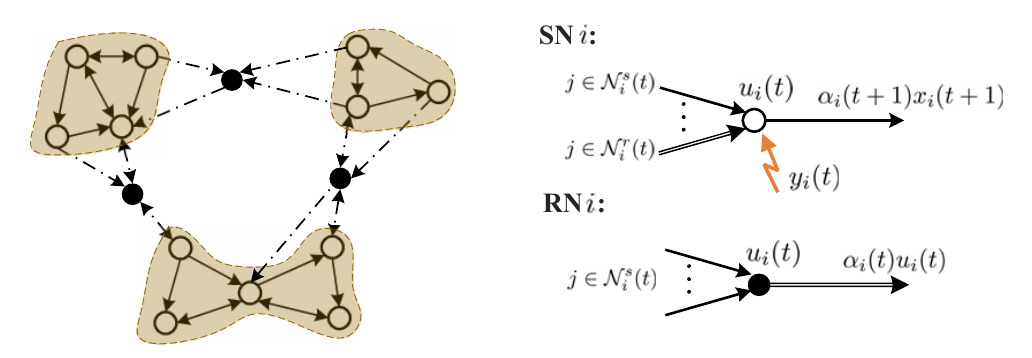}
 \caption{(left) A relay-assisted sensor network. (right) A schematic view of node operation at SNs and RNs.}\label{fig:example}
\end{figure}

   Consider a relay-assisted sensor network consisting of SNs and RNs that monitors a field of interest (see Fig.~\ref{fig:example} (left)). 
   There are  $N$ nodes observing a common phenomenon with an unknown parameter $\theta\in \mathbb{R}$.  Without loss of generality, we label the SNs from 1 to $M$, and RNs from $M+1$ to $N$. We consider the situation that on-board sensors of SNs are subject to sensor faults from time to time.  Thus SN $i$ can occasionally measure the ambient parameter $\theta$  and obtain a noisy version
\begin{equation}\label{eq:measurement}
  y_i(t)=\theta+w_i(t), \ i=1,2,\dots, M,
\end{equation}
  where $w_i(t)$ is the measurement noise with zero mean. 

  Due to the intermittent nature of SNs, cooperation  between the nodes is necessary for reliable estimation of  $\theta$. 
  To this end, nodes exchange information with their neighboring nodes.
  such that the measurements at SNs can propagate through the whole network. 
   During the information exchange, we do not consider  transmission delays, but focus on the scenario where transmissions suffer from channel fading and additive noise.

  \emph{Graph Description:} We model the communication topology over which the nodes exchange information as a \emph{dynamic directed graph} $\mathcal G(t)=(\mathcal{V},\mathcal{E}(t))$,  where $\mathcal{V}=\mathcal{I}_s\cup \mathcal{I}_r$  is the set of the nodes with $\mathcal{I}_s=\{1,2,\ldots,M\}$, $\mathcal{I}_r=\{M+1,M+2,\ldots,N\}$ being the sets of SNs and RNs, respectively, and $\mathcal{E}(t)\subset \mathcal{V}\times \mathcal{V}$ is the set of communication links at time $t$. The structure of the graph is described by the 0-1 adjacency matrix $\mathbf{A}(t)=[a_{ij}(t)]_{N\times N}$, where $a_{ij}(t)=1 \Leftrightarrow (j,i)\in \mathcal{E}(t)$. We do not allow self-loops, i.e., $a_{ii}(t)=0$, $\forall i$. The set of \emph{neighbors} of node $i$ is defined as $\mathcal{N}_i(t) =\left\{j: a_{ij}(t)>0\right\}$ with $|\mathcal{N}_i(t)|$ being its cardinality. $\mathcal{N}_i^s(t)\triangleq \mathcal{N}_i(t)\cap \mathcal{I}_s$ and $\mathcal{N}_i^r(t)\triangleq \mathcal{N}_i(t)\cap \mathcal{I}_r$ stand for the SN and RN neighbors of node $i$, respectively. 

  \emph{Multi-access Scheme with Fading:}  Let $x_i(t)$ be the estimate of node $i$ at time $t$.  With the analog-and-forward scheme, node $i$ scales the data using the amplification factor $\alpha_{i}(t)>0$ and then forwards it to its neighbors over fading channels. 
  The received data at node $i$ can then be expressed as
\begin{equation}\label{eq:multiaccess}
 u_i(t) \triangleq
 \sum\limits_{j\in\mathcal{N}_i(t)}\bigl(\alpha_{j}(t) h_{ij}(t) x_j(t)+v_{ij}(t)\bigr), \ \forall i\in \mathcal{V},
\end{equation}
  where $h_{ij}(t)$ is the random fading coefficient of the link from node $j$ to node $i$, representing the real-valued envelope of the complex channel gain, and $v_{ij}(t)$ is the receiver noise in the transmission from node $j$ to node $i$.  

  We take into account channel fading by incorporating the fading coefficient into the entries of the adjacent matrix $\mathbf{A}(t)$ as $\mathbf{A}(t)=[a_{ij}(t)h_{ij}(t)]_{N\times N}$.  To facilitate our ensuing analysis, we make the following standard assumptions:

  \emph{Assumption 1:} There exists a constant $B>0$ such that the union graph $\mathcal{G}_{[k B, (k+1) B)}=\bigl(\mathcal{V}, \bigcup_{k B}^{(k+1)B-1} \mathcal{E}(t)\bigr)$  is strongly connected on \emph{average}\footnote{This means that the adjacent matrix $\sum_{t=k B}^{(k+1)B-1}\mathbb{E}\{\mathbf{A}(t)\}$ corresponds to a strongly connected graph, where the expectation is taken over all realizations of the fading coefficients $\{h_{ij}(t)\}_{i,j}$.} for all $k\geq 0$, i.e., there exists a multi-hop path in the union graph connecting any two vertices.

  \emph{Assumption 2:} At least one SN has access to a measurement every $T>0$ time steps. The measurement noise  sequence $\bigl\{\mathbf{w}(t)=[w_1(t),\dots,w_M(t)]^T\bigr\}_{t\geq 0}$ is i.i.d. zero mean with finite variance.
  
  \emph{Assumption 3:}  The fading coefficients $\{h_{ij}(t)\}_{i,j}$ have $\bar{h}_{ij}\triangleq\mathbb{E}\{h_{ij}(t)\}>0$ and finite variances, and are independent across transmitters and  time steps. The receiver noise $\{v_{ij}(t)\}_{i,j}$ are zero-mean random variables with finite variance, and are temporally independent. 

 \emph{Assumption 4:} All $w_i(t)$, $v_{ij}(t)$ and $h_{ij}(t)$ are mutually independent with respect to $t$.  

\begin{remark}\label{rem:assumptions}
  Assumption 1 is a joint connectivity condition, which ensures the repeated influence of the nodes on each other. 
  Such kind of condition is widely used for distributed algorithms with time-varying topologies, e.g., \cite{Huang12,TsitBertAtha86,XiaoBoyLal05,ZhangZhang12}. It is important to remark that, we only require the union graph to be strongly connected on average. In fact, it is possible to have all these instantiations $\mathcal{G}(t)$ and $\mathcal{G}_{[k B, (k+1) B)}$ to be disconnected. 
  Assumption~2 requires a uniform measurement speed of SNs to guarantee rich information of the unknown parameter. Note that the measurement noises  at different SNs may be correlated. Assumption 2 only states temporal independence. Assumption~3 is reasonable when the nodes are dispersed or the coherence time is small compared with the duration of one round of transmissions \cite{ChanSwamZhaoScag10,MostMalm10}.
\end{remark}

\section{Distributed innovation-based estimation algorithm}\label{sec:algorithm}
  As in standard point estimation theory, least-squares estimator is one widely used estimator, which can be recursively updated by adding a refinement
  term to the previous estimate \cite{BarLiKir04}.  The refinement term is usually referred to as the innovation. This basic idea has been adopted for several distributed algorithms, e.g., \cite{KarMouRam12,StanSti11,ZhangZhang12}. 

  In this paper, we also adopt this idea to propose our algorithm. Each SN first takes measurement, then combines it with data from its neighbors to update its prior estimate. At each SN $i$, besides the occasional measurement $y_i(t)$, the data $u_{i}(t)$ from its neighbors can be regarded as another source of measurement. In this way, we can construct two terms of innovations: own measurement innovation and innovation of local data from neighbors to refine the previous estimate. To be specific,  if SN $i$ has a measurement, then it will do the following update: 
\begin{equation}\label{eq:stamodeI}
  x_i(t+1)=x_i(t)+a(t)[y_i(t)-x_i(t)]
  +a(t)\left[ u_{i}(t)-b_i(t) x_i(t)\right], \ i \in \mathcal{I}_s,
\end{equation}
  where $a(t)>0$ and $b_i(t)$ are appropriately chosen weights, which will be discussed in Section~\ref{sec:algorithmdesign}.
  On the other hand, if SN $i$ suffers from sensor failure, then it only uses $u_i(t)$ to calibrate the previous estimate 
\begin{equation}\label{eq:stamodeI'}
  x_i(t+1)=x_i(t)+a(t)[u_{i}(t)- b_i(t) x_i(t)], \ i \in \mathcal{I}_s.
\end{equation}
  Note that in \eqref{eq:stamodeI} and \eqref{eq:stamodeI'}, we let $b_i(t)=0$ if $\mathcal{N}_i(t)=\emptyset$.

  As for RNs, there is no further processing of the incoming data. With an abuse of notation, we denote the reception at each RN $i$ as its estimate, namely,
  \begin{equation}\label{eq:stamodeII}
  x_i(t)=u_i(t), \ i \in \mathcal{I}_r.
\end{equation}

  The recursions in \eqref{eq:stamodeI}-\eqref{eq:stamodeII} constitute our distributed estimation algorithm. The difference between node operations at SNs and RNs is schematically illustrated in Fig.~\ref{fig:example} (right).  We note that there are two kinds of weights $a(t)$ and $\{b_i(t)\}_{i}$ introduced in the algorithm \eqref{eq:stamodeI} and \eqref{eq:stamodeI'} for SNs. Appropriate forms of $\{b_i(t)\}_{i}$ are crucial in attenuating the effects of the channel fading.  This formulation differs from  those estimation algorithms proposed for sensor networks without taking the effect of channel fading into account\cite{KarMouRam12,StanSti11,ZhangZhang12}. 

\begin{table*}[!t]
  \caption{Coefficient matrices of system \eqref{eq:system}}
  \label{tab:notation}
  \begin{center}
    \begin{tabular}{c||l}
    \hline
    Notation & \hspace{0.25\linewidth} Description \\
    \hline
    $\mathbf{\Delta}(t)=\text{diag}\{\chi_1(t),\chi_2(t),\dots, \chi_M(t)\}$ & $\chi_i(t)=1$ or 0 indicating whether SN $i$ has a measurement at time $t$ or not \\
    $\mathbf{L}(t)=[l_{ij}(t)]_{M\times M}$ &     $l_{ij}(t)=\begin{cases}
  -\alpha_{j}(t)\Bigl[h_{ij}(t)+\sum_{\{k: k\in\mathcal{N}_i^r(t), j\in \mathcal{N}_k^s(t)\}}\alpha_k(t) h_{ik}(t) h_{kj}(t)\Bigr],& j\in\mathcal{N}_i^s(t),\\
  -\alpha_{j}(t)\sum_{\{k: k\in\mathcal{N}_i^r(t), j\in \mathcal{N}_k^s(t)\}}\alpha_k(t) h_{ik}(t) h_{kj}(t),& j\not\in\mathcal{N}_i^s(t)\cup \{i\},\\
  -\sum_{j\neq i} l_{ij}(t), &j=i
  \end{cases}$\\   
  $\mathbf{\Gamma}(t)=\text{diag}\{\gamma_1(t),\gamma_2(t),\dots,\gamma_M(t)\}$ & 
  $\gamma_i(t)=  \sum_{j\in \mathcal{N}_i^s(t)} \alpha_j(t) h_{ij}(t)+\sum_{j\in \mathcal{N}_i^r(t)} \alpha_j(t) h_{ij}(t)\sum_{k\in \mathcal{N}_j^s(t)} \alpha_k(t) h_{jk}(t)-b_i(t)$\\     
    \hline
    \end{tabular}
  \end{center}
\end{table*}

  In order to simplify the presentation, we consider the case that each SN can communicate with other SNs directly or via one-hop RN\footnote{For the case of multi-hop RNs, we can use the method of graph transformation proposed in our previous work \cite{ZhuChenMaYangGuan15} to construct an equivalent network consisting of only one-hop RNs.}. Moreover, we assume that once the data reach one RN, it can be forwarded to its neighbors instantaneously\footnote{Our results can be easily extended to the case that RNs should wait for some time before forwarding their received data. In this case, outdated information of RNs will be transmitted to the neighbors. This leads to a time-delay algorithm. The method developed in this paper is still applicable by transforming it to a delay-free system as in\cite{Huang12,LiuXieZhang11}. The only difference is that we should deal with a higher dimensional system.}. With this setting, we have $\mathcal{N}_i(t)=\mathcal{N}_i^s(t)$, for all RNs $i\in \mathcal{I}_r$. Then substituting \eqref{eq:multiaccess} into \eqref{eq:stamodeII}, we have 
\[
  x_i(t)=\sum_{j\in \mathcal{N}_i^s(t)} \alpha_{j}(t) h_{ij}(t) x_j(t), \ i \in \mathcal{I}_r,
\]
  which along with \eqref{eq:multiaccess} implies that for SN $i\in \mathcal{I}_s$,
\setlength\multlinegap{0pt}
\begin{equation}\label{eq:receivedSN}
u_i(t)=\sum_{j\in \mathcal{N}_i^s(t)} \alpha_{j}(t) h_{ij}(t) x_j(t) +\sum_{j\in \mathcal{N}_i^r(t)} \alpha_j(t)h_{ij}(t) \sum_{k\in \mathcal{N}_j^s(t)} \alpha_{k}(t) h_{jk}(t) x_k(t) + v_i(t),
\end{equation}
where the quantity $v_i(t)$ collects all the receiver noises at SN~$i$,
\[
 v_i(t)=\sum_{j\in \mathcal{N}_i(t)} v_{ij}(t)+\sum_{j\in \mathcal{N}_i^r(t)} \alpha_j(t)h_{ij}(t) \sum_{k\in \mathcal{N}_j^s(t)} v_{jk}(t).
\]

 Let $\mathbf{x}(t)\!=[x_1(t),\ldots,x_{M}(t)]^T$, $\mathbf{y}(t)\!=[y_1(t),\ldots,y_{M}(t)]^T$ and $\mathbf{v}(t)=[v_1(t),\dots,v_M(t)]^T$. Using matrix-vector notation, we can 
 reformulate \eqref{eq:stamodeI}, \eqref{eq:stamodeI'} and \eqref{eq:receivedSN} more compactly into
\begin{equation}\label{eq:system}
  \mathbf{x}(t+1)=\bigl(\mathbf{I}-a(t)\mathbf{\Phi}(t)\bigr) \mathbf{x}(t)+a(t)\mathbf{\Delta}(t) \mathbf{y}(t)+a(t)\mathbf{v}(t),
\end{equation}
 where $\mathbf{\Phi}(t)=\mathbf{\Delta}(t)+\mathbf{L}(t)-\mathbf{\Gamma}(t)$, $\mathbf{\Delta}(t)$ is the diagonal indicator matrix, $\mathbf{L}(t)$ is a random matrix and $\mathbf{\Gamma}(t)$ is the random uncertainty matrix (see Table~\ref{tab:notation}). 

 Some basic properties of \eqref{eq:system} that will be frequently used in the sequel  are stated in the next result.

\begin{proposition}\label{pro:propertyL}
 Consider the system \eqref{eq:system}. Suppose that Assumptions 2-4 hold, then we have the following properties:
 \begin{enumerate}
 \item[i)] $\{\mathbf{x}(t), \mathcal{F}_t\}_{t\geq 0}$ is a Markov process, where $\mathcal{F}_t=\sigma\bigl(\mathbf{w}(s), \mathbf{v}(s), \mathbf{L}(s), \mathbf{\Gamma}(s), 0\leq s<t\bigr)$ is the filtration.
  \item[ii)] $\mathbf{L}(t)\mathbf{1}=\mathbf{0}$ with $\mathbb{E}\{l_{ij}(t)\}\leq 0$, $\forall j\neq i$. If, further, $(j,i)\in \mathcal{E}(t)$ or $(j,p), (p,i)\in \mathcal{E}(t)$ for some $p\in \mathcal{I}_r$, then $\mathbb{E}\{l_{ij}(t)\}\leq-\alpha_j(t)\min\{\bar{h}_{ij}, \alpha_q(t)\bar{h}_{ip}\bar{h}_{pj}\}$.
  \end{enumerate} 
 \end{proposition}
 \begin{IEEEproof}
  See Appendix~\ref{app:proofpropertyL}.
 \end{IEEEproof}

\section{Asymptotic properties of proposed algorithm}\label{sec:convergence}
  In this section, we will examine two prevailing asymptotic properties of the proposed algorithm \eqref{eq:system} with respect to the first and second moments \cite{BarLiKir04,PinSch01}, respectively. For the first moment, it is related to asymptotic unbiasedness, while for the second moment, what matters is asymptotic consistency.

  To this end,  we let $e_i(t)\triangleq x_i(t)-\theta$ be the estimation error for SN $i\in\mathcal{I}_s$, and $\mathbf{e}(t)=[e_1(t),\ldots,e_M(t)]^T$ be the error vector.  Subtracting $\theta\mathbf{1}$ from both sides of \eqref{eq:system} and using \eqref{eq:measurement}, one then obtains the error dynamics
\begin{equation}\label{eq:errdyn}
 \mathbf{e}(t+1)=\bigl(\mathbf{I}-a(t)\mathbf{\Phi}(t)\bigr)\mathbf{e}(t)
 +a(t) (\theta \mathbf{\Gamma}(t)\mathbf{1}+\mathbf{\Delta}(t) \mathbf{w}(t)+\mathbf{v}(t)),
\end{equation}
 where we have used the property that $\mathbf{L}(t)\mathbf{1}=\mathbf{0}$ (see Proposition~\ref{pro:propertyL}).

\subsection{Change of scale of time steps}
 Let $\tau\triangleq \max\{T,B\}$, $t_k\triangleq k\tau$ and define the transition matrix $\mathbf{\Psi}_{t_{k+1},t_k}\triangleq \prod_{s=t_k}^{t_{k+1}-1} (\mathbf{I}-a(s)\mathbf{\Phi}(s))$. Continuing the recursion of \eqref{eq:errdyn}, we can obtain
 \setlength\multlinegap{0pt}
\begin{equation}\label{eq:errordynnewscale}
\mathbf{e}(t_{k+1})=\mathbf{\Psi}_{t_{k+1},t_k} \mathbf{e}(t_k)+\sum_{s=t_k}^{t_{k+1}-1} a(s)\mathbf{\Psi}_{t_{k+1},s+1} \bigl(\theta \mathbf{\Gamma}(s)\mathbf{1}+\mathbf{\Delta}(s) \mathbf{w}(s)+\mathbf{v}(s)\bigr).
\end{equation}

 By Assumptions 1 and 2, the union graph $\mathcal{G}_{[t_k,t_{k+1})}$ is strongly connected on average, and there exists at least one $t_k\leq s_k<t_{k+1}$ such that $\sum_{i=1}^M \chi_i(s_k)\geq 1$. 
 we have the following important property of the error dynamics \eqref{eq:errordynnewscale} in the new time scale.
 
 \begin{proposition}\label{pro:irredu}
 Suppose that $\mathbb{E}\{\mathbf{\Gamma}(t)\}=\mathbf{0}$ and $a(t)$ can be sufficiently small, then under Assumptions 1 and 2, the dominant term $\mathbf{I}-\sum_{s=t_k}^{t_{k+1}-1}a(s)\mathbb{E}\{\mathbf{\Phi}(s)\}$ of $\mathbb{E}\{\mathbf{\Psi}_{t_{k+1},t_k}\}$ corresponds to an irreducible matrix for all $k=0,1,\ldots$, and at least one row has sum strictly less than 1.
 \end{proposition}
 \begin{IEEEproof}
 See Appendix~\ref{app:proofirredu}.
\end{IEEEproof}

The above result will show its advantage of the change of scale from 1 to $\tau$ in the consistency analysis, where some nice properties of irreducible matrices will be explored.

 \subsection{First moment analysis: Asymptotic unbiasedness}
 Denote $\bar{\mathbf{e}}(t)\triangleq \mathbb{E}\{\mathbf{e}(t)\}$, $\bar{\mathbf{L}}(t)=[\bar{l}_{ij}(t)]_{M\times M}\triangleq \mathbb{E}\{\mathbf{L}(t)\}$ and $\bar{\mathbf{\Phi}}(t)\triangleq \mathbb{E}\{\mathbf{\Phi}(t)\}$. Taking expectations of both sides of \eqref{eq:errdyn} and recalling the zero-mean assumptions of $\mathbf{w}(t)$ and $\mathbf{v}(t)$, one can obtain
\begin{equation}\label{eq:1stmoment}
  \bar{\mathbf{e}}(t+1)=\bigl(\mathbf{I}-a(t)\bar{\mathbf{\Phi}}(t)\bigr)\bar{\mathbf{e}}(t)+a(t) \theta \mathbb{E}\{\mathbf{\Gamma}(t)\}\mathbf{1}.
\end{equation}

  Asymptotic unbiasedness requires that $\bar{\mathbf{e}}(t)\to 0$ as $t\to \infty$. Lyapunov method is widely used to investigate the asymptotic properties of dynamical systems. However, as far as the system \eqref{eq:1stmoment} is concerned, it is noted that $\bar{\mathbf{\Phi}}(t)$ is time-varying and nonsymmetric, and thus a common Lyapunov function, which is essential for the time-invariant or symmetric cases \cite{KarMou11,KarMouRam12,ZhangZhang12,ZhuChenMaYangGuan15}, is generally not available. Furthermore, the eigenvalues of $\mathbf{I}-\bar{\mathbf{\Phi}}(t)$ can be sufficiently close to 1, if we use decaying weight that $\lim_{t \to \infty} a(t)=0$. In this case, \eqref{eq:1stmoment} is a degenerated linear system.
  These two aspects render the classical results of linear systems in \cite{Rugh96} not suitable for use here. To address this issue, we utilize a different technique, which  allows us to deal with the general case of asymmetric communications in a tractable framework.

   For this purpose, we reorder the states $\{\bar{e}_1(t),\dots,\bar{e}_M(t)\}$ in such a way  that SN $i_t$ has the $i$-th largest value among all SNs at time $t$, namely, $\bar{e}_{1_t}(t)\geq\dots \geq \bar{e}_{M_t}(t)$, where $\{1_t,2_t,\dots,M_t\}$ is a permutation of $\{1,2,\dots,M\}$. This operation will enable $\bar{e}_{i_t}(t)$ to behave in a  more orderly manner, which is also used for consensus algorithms with time-varying graphs \cite{Huang12}.
   For clarity of presentation, we denote
\begin{equation}\label{eq:defz}
  z_i(t)\triangleq \bar{e}_{i_t}(t),\  \forall 1\leq i\leq M.
\end{equation}
  According to the ordering of $\bar{e}_{i_t}$, we have $z_1(t)\geq z_2(t)\geq \dots\geq z_M(t)$, $\forall t\geq 0$.

  We shall establish the asymptotic properties of \eqref{eq:1stmoment} via the ordered sequence  $\{z_i(t)\}_{1\leq i\leq M, t\geq 0}$. We first give some important properties of the ordered sequence.

 \begin{lemma}\label{lem:propertiesofz}
   Consider the sequences $\{z_i(t)\}_{1\leq i\leq M, t\geq 0}$ governed by \eqref{eq:1stmoment} and \eqref{eq:defz}. Suppose that $\mathbb{E}\{\mathbf{\Gamma}(t)\}=\mathbf{0}$, and there exists a time $t^*>0$ such that 
\begin{equation}\label{eq:condition_at}
   a(t)\bigl(1+\max_i \bar{l}_{ii}(t)\bigr)\leq 1, \ t\geq t^*, 
\end{equation}
   then 
\begin{enumerate}
   \item[i)] there is a constant $c_0>0$ so that $\max_i |z_i(t)|\leq c_0$, $\forall t$;
  \item[ii)] we have either $\lim_{t\to\infty} z_M(t)=z_{M\infty}\leq 0$ or $\lim_{t\to\infty} z_1(t)=z_{1\infty}\geq  0$, where $z_{M\infty}$ and $z_{1\infty}$ are finite limits, respectively.
\end{enumerate}
 \end{lemma}

   The proof is given in Appendix~\ref{app:proofpropertiesofz}. An outline is as follows: Statement i) is due to the fact that $\|\bar{\mathbf{e}}(t)\|_{\infty}$ is nonincreasing on the interval $[t^*,\infty)$. We prove ii) by exploiting the properties of $\bar{\mathbf{L}}(t)$ in Proposition~\ref{pro:propertyL} to show that either $z_1(t)$ is nonincreasing or $z_M(t)$ is nondecreasing for large $t$.

   Our intuition about reordering of the states $\bar{e}_i(t)$, $1\leq i\leq M$, is substantiated by Lemma~\ref{lem:propertiesofz}(ii), which implies that the behavior of the largest or smallest state does follow a very orderly manner. With this nice property, we are expected to go further by imposing more conditions on \eqref{eq:system}. This is formally stated in the next theorem, which establishes asymptotic unbiasedness of \eqref{eq:system}. 
\begin{theorem}\label{thm:asyunb}
Consider the estimation algorithm given in the form of \eqref{eq:system} with bounded amplification factors $\{\alpha_i(t)\}_{i=1}^N$.  Assume that $\mathbb{E}\{\mathbf{\Gamma}(t)\}=\mathbf{0}$, and there is a  constant $0< \underline{c}<1$ such that
\begin{equation}\label{eq:persistence}
 \underline{c} \leq \frac{a(t+1)}{a(t)}\leq 1, \ \text{for large} \ t, \  \text{and} \  \lim_{t\to\infty}a(t)= 0.  
\end{equation}
If further, for any $\nu>0$, there is a constant $\kappa>0$ such that 
\begin{equation}\label{eq:persistence2}
  \exists T(t,\nu)>0, \ \kappa\nu \leq \sum_{s=t}^{t+T(t,\nu)} a(s)\leq \nu, \ \text{for large} \ t,
\end{equation}
then under Assumptions 1-3, the estimate sequence $\{\mathbf{x}(t)\}_{t\geq 0}$ is asymptotically unbiased, i.e., 
\[
\lim_{t\to \infty}\mathbb{E}\{\mathbf{x}(t)\}=\theta \mathbf{1}.
\]
\end{theorem}

  See Appendix~\ref{app:proofunbiasedness} for the proof.
  A sketch is as follows. We proceed in two steps. In step~1, we extend Lemma~\ref{lem:propertiesofz} to networks with joint connectivity. The procedure is directly operated on the process \eqref{eq:errordynnewscale} in the new time scale.
  Specifically, we  show that all the ordered sequences $\{z_i(t)\}_{t\geq 0}$ are convergent and have the same limit $z_{\infty}$. Consequently, we have $\lim_{t\to\infty} \bar{\mathbf{e}}(t)=z_{\infty}\mathbf{1}$.  In step 2, using the method of recurrent inequalities, we argue that there is a subsequence $\{\|\bar{\mathbf{e}}(s_k)\|_1\}_{k\geq 1}$ of $\{\|\bar{\mathbf{e}}(t)\|_1\}_{t\geq 0}$ converging to 0. Then we conclude that $z_{\infty}=0$, achieving asymptotic unbiasedness.

\subsection{Second moment analysis: Asymptotic consistency}
 Now we turn to the second moment analysis of the proposed algorithm. Specifically, we will investigate the convergence of $\mathbb{E}\{\|\mathbf{e}(t)\|_2^2\}=\mbox{trace}(\mathbb{E}\{\mathbf{e}(t)\mathbf{e}^T(t)\})$.  
  Denote by $V(t)=\|\mathbf{e}(t)\|_2^2$, then 
 using the Markov property of \eqref{eq:system} (see Proposition~\ref{pro:propertyL}), we can derive from \eqref{eq:errdyn} that
  \begin{align}\label{eq:2nditeration}
   \mathbb{E}\{V(t+1)\}&=\mathbb{E}\bigl\{\|(\mathbf{I}-a(t)\mathbf{\Phi}(t))\mathbf{e}(t)\|_2^2\bigr\}+2\theta a(t) \bar{\mathbf{e}}^T(t)\mathbb{E}\bigl\{(\mathbf{I}-a(t) \mathbf{\Phi}(t))^T \mathbf{\Gamma}(t)\bigr\}\mathbf{1} \notag\\
   &\relphantom{=}{}+a^2(t)\mathbb{E}\{\|\theta \mathbf{\Gamma}(t)\mathbf{1}+\mathbf{\Delta}(t) \mathbf{w}(t)+\mathbf{v}(t)\|_2^2\}.
  \end{align}

  Asymptotic consistency means that $\mathbb{E}\{\|\mathbf{e}(t)\|_2^2\}\to 0$ as $t\to \infty$. We know that consistency implies asymptotic unbiasedness \cite[p.456]{PinSch01}. Hence, it is natural that more conditions are needed for the consistency analysis. As a motivation, let us first examine two simple cases:
\begin{itemize}
  \item $\chi_i(t)\equiv 0$, $\forall i$, and $\mathcal{E}(t)\equiv\emptyset$. That is, there are no measurements and communications between nodes.

  In this case, it is clear that \eqref{eq:2nditeration} becomes $\mathbb{E}\{V(t+1)\}=\mathbb{E}\{V(t)\}=\mathbb{E}\{V(0)\}$. Consistency can not be achieved!  
  \item there is a SN with $\chi_{i_0}(t)=1$ occasionally  and $\mathcal{E}(t) \equiv\emptyset$. 
  That is, some SNs have measurements occasionally, but there are no communications betweens nodes.  

  In this case, we can obtain from \eqref{eq:2nditeration} that 
\begin{equation}\label{eq:2nditeration_simple}
 \mathbb{E}\{V(t+1)\}\leq  \mathbb{E}\{V(t)\} 
 -2 \zeta a(t) \chi_{i_0}(t) \mathbb{E}\{e_{i_0}(t)^2\}
 +a^2(t) \mathbb{E}\{\|\mathbf{w}(t)\|_2^2\},
\end{equation}
 where $\zeta>0$ is a constant, and we use the fact that $a(t)\leq 1-\zeta$, $\forall t$, in view of \eqref{eq:persistence}. In general, the iteration \eqref{eq:2nditeration_simple} does not converge. One widely used condition to enable convergence is square-summability of $a(t)$ \cite{Huang12,KarMouRam12,StanSti11,TsitBertAtha86}, i.e.,
\begin{equation}\label{eq:conditionconsis}
 \sum_{t=0}^{\infty} a^2(t)<\infty.
\end{equation} 
 With this condition, the Robbin-Siegmund theorem \cite[p.50]{Pol87} implies the existence of $\lim_{t\to \infty}\mathbb{E}\{V(t)\}$. Consistency can be guaranteed if further the second term of the RHS of \eqref{eq:2nditeration_simple} is of the order $\mathcal{O}(-a(t) V(t))$. This is possible by considering \eqref{eq:errordynnewscale} in the new time scale, once we recall the properties that $\sum_{i=1}^M\sum_{s=t_k}^{t_{k+1}-1} \chi_i(t)\geq 1$ and graph $\mathcal{G}(t)$ is jointly strongly connected on average by Assumption~1. 
\end{itemize}

 The above discussions inspire our second main result regarding asymptotic consistency of \eqref{eq:system}. 
\begin{theorem}\label{thm:consistency}
Consider the estimation algorithm given in the form of \eqref{eq:system} with bounded amplification factors $\{\alpha_i(t)\}_{i=1}^N$ and weights $\{b_i(t)\}_{i=1}^M$.  Assume that $\mathbb{E}\{\mathbf{\Gamma}(t)\}=\mathbf{0}$, $a(t)$ satisfies conditions \eqref{eq:persistence},  \eqref{eq:persistence2} and \eqref{eq:conditionconsis}. Moreover, there is a constant $c^*>0$ such that the minimal eigenvalue
\begin{equation}\label{eq:boundcondconsis}
\lambda_{\min}(\mathbf{J}_k)\geq c^*a(t_k),
\end{equation}
where $\mathbf{J}_k=\sum_{s=t_k}^{t_{k+1}-1}a(s)(2\mathbf{\Delta}(s)+\bar{\mathbf{L}}(s)+\bar{\mathbf{L}}^T(s))$.
Then under Assumptions 1-4, the estimate sequence $\{\mathbf{x}(t)\}_{t\geq 0}$ is asymptotically consistent, i.e.,  
\[
 \lim_{t\to\infty} \mathbb{E}\{\|\mathbf{x}(t)-\theta \mathbf{1}\|_2^2\}=0.
\]
\end{theorem}

 The proof is given in Appendix~\ref{app:proofconsistency}.
 The basic idea is to first investigate the convergence of $\mathbb{E}\{V(t_{k})\}$ by considering \eqref{eq:errordynnewscale} in the new time scale. The convergence is guaranteed by the Robbin-Siegmund theorem on random sequences. Secondly, we need to pass from the convergence of the subsequence $\{\mathbb{E}\{V(t_k)\}\}_{k\geq 0}$ to the convergence of $\{\mathbb{E}\{V(t)\}\}_{t\geq 0}$, and draw the conclusion on asymptotic consistency.

\subsection{Discussion on consistency }
 Compared with the first moment analysis in Theorem~\ref{thm:asyunb}, we impose two more constraints in examining the second moment: one is the square-summability condition \eqref{eq:conditionconsis},  and the other is condition \eqref{eq:boundcondconsis}. The former is introduced to suppress the involved noises. The latter specifies a requirement on the asymmetric topology so that the mirror graph related with $\bar{\mathbf{L}}(t)+\bar{\mathbf{L}}^T(t)$ has some desired properties.

 In the following, we will give some discussions on the condition \eqref{eq:boundcondconsis}. 
\subsubsection{Graphs with asymmetric links}  
 We remark that  condition \eqref{eq:boundcondconsis} is not needed if graph $\mathcal{G}$ is fixed, and $\bar{\mathbf{L}}(t)\equiv \bar{\mathbf{L}}$ is time-invariant. In fact, there is a positive vector $\boldsymbol{\omega}$ such that $\boldsymbol{\omega}^T\bar{\mathbf{L}}=\mathbf{0}$ for connected graph $\mathcal{G}$ \cite{ZhuChenMaYangGuan15}. We can then use $\mathbf{e}^T(t) \mathbf{\Xi} \mathbf{e}(t)$ instead of $\|\mathbf{e}(t)\|_2^2$ as the Lyapunov candidate in the proof of Theorem~\ref{thm:consistency}, where $\mathbf{\Xi}=\text{diag}\{\boldsymbol{\omega}\}$. In this case, we can show that \emph{strong consistency} \cite[p.455]{PinSch01}, i.e., $\mathbb{P}\{\lim_{t\to \infty}\|\mathbf{x}(t)-\theta \mathbf{1}\|=0\}=1$, can be guaranteed using the similar arguments as in \cite{ZhuChenMaYangGuan15}.

  For the time-varying graph $\mathcal{G}(t)$, we rewrite $\mathbf{J}_k$ as 
\[
  \mathbf{J}_k=2\sum_{s=t_k}^{t_{k+1}-1}a(s)\bigl( \max_{i} \bar{l}_{ii}(s)+1\bigr) \mathbf{I}-\mathbf{\Omega}_k,
\]
  where $\mathbf{\Omega}_k=\sum_{s=t_k}^{t_{k+1}-1}a(s)[2(\max_{i} \bar{l}_{ii}(s)+1) \mathbf{I}-\mathbf{\Delta}(s)-\mathbf{\bar{L}}(s)-\mathbf{\bar{L}}^T(s)]$.   Clearly, $\mathbf{\Omega}_{k}$ is a nonnegative matrix. Furthermore, by Proposition~\ref{pro:irredu}, we know that $\mathbf{\Omega}_{k}$ is irreducible. Hence, from Perron-Frobenius theory \cite[p.11]{Minc88}, $\rho(\mathbf{\Omega}_{k})$ is the maximal eigenvalue of $\mathbf{\Omega}_{k}$. It thus follows that 
\[
  \lambda_{\min}(\mathbf{J}_k)=2\sum_{s=t_k}^{t_{k+1}-1}a(s) \bigl(\max_{i} \bar{l}_{ii}(s)+1\bigr)\mathbf{I}- \rho(\mathbf{\Omega}_{k}).
\]
 In order to enforce \eqref{eq:boundcondconsis}, it suffices to establish some upper bounds of  $\rho(\mathbf{\Omega}_{k})$. One well-known upper bound is $r_{\max}(\mathbf{\Omega}_{k})$ \cite[p.24]{Minc88}. However, it is too conservative for certain types of topologies, e.g., $\mathbf{1}^T\bar{\mathbf{L}}(t)=\bar{\mathbf{L}}(t) \mathbf{1} \equiv\mathbf{0}$. We can only get $\lambda_{\min}(\mathbf{J}_k)\geq 0$ in this case, which violates \eqref{eq:boundcondconsis}.
 
 Next, we will give a tighter and easy-to-check bound for irreducible nonnegative matrices using generalized Perron complement \cite{LuNg04}. We first introduce some notations.
 Let $\mathbf{B}_{S_1,S_2}$ be the submatrix of $\mathbf{B}\in \mathbb{R}^{M\times M}$ whose rows and columns are indexed by nonempty sets $S_1,S_2\subset \mathcal{V}$. If $S_1=S_2$, we simply denote it as $\mathbf{B}_{S_1}$. For any set $S\subset \mathcal{V}$ and its complement $S^c=\mathcal{V}\backslash S$, define 
\[
  \iota(\mathbf{B})\triangleq \frac{1}{2}\max_{i,j}\Bigl(r_{ij}^{+}+\sqrt{(r_{ij}^{-})^2+4r_{ij}}\Bigr),
\]
 where $r_{ij}^{+}=r_i(\mathbf{B}_{S})+r_{j}(\mathbf{B}_{S^c})$, $r_{ij}^{-}=r_i(\mathbf{B}_{S})-r_{j}(\mathbf{B}_{S^c})$ and $r_{ij}=r_i(\mathbf{B}_{S,S^c})r_j(\mathbf{B}_{S^c,S})$. 
\begin{proposition}\label{pro:boundmaxeigenvalue}
  For any irreducible nonnegative matrix $\mathbf{B}$,  we have
  \[
  \rho(\mathbf{B})\leq \iota(\mathbf{B})\leq r_{\max}(\mathbf{B}).
  \]
\end{proposition}
\begin{IEEEproof}
See Appendix~\ref{app:proofboundmaxeigenvalue}.
\end{IEEEproof}

Clearly, the upper bound $\iota(\mathbf{B})$ in the proposition improves the bound $r_{\max}(\mathbf{B})$. It thus follows that condition \eqref{eq:boundcondconsis} is satisfied if 
\begin{equation}\label{eq:boundcondconsis2}
  \iota(\mathbf{\Omega}_k)\leq 2\sum_{s=t_k}^{t_{k+1}-1}a(s)\bigl(\max_{i} \bar{l}_{ii}(s)+1\bigr)-c^*a(t_k).
\end{equation}
 holds. We emphasize that the RHS of \eqref{eq:boundcondconsis2} is positive for small $c^*$. To see this, we use \eqref{eq:persistence} to get $a(s)/a(t_k)\geq \underline{c}^{s-t_k}$, $\forall s\geq t_k$, which further gives
\begin{equation}
 \text{RHS}\geq a(t_k) \sum_{s=t_k}^{t_{k+1}-1} \Bigl(2 \underline{c}^{s-t_k}\max_{i} \bar{l}_{ii}(s)+\frac{1-\underline{c}^{\tau}}{1-\underline{c}}-c^*\Bigr)>0,
\end{equation}
 provided that  $c^*\leq (1-\underline{c}^{\tau})/(1-\underline{c})$.

\subsubsection{Graphs with symmetric structures}
  For the communication topology $\mathcal{G}(t)$, if some symmetric structures, e.g., matrix $\bar{\mathbf{L}}(t)$ is symmetric (i.e., $\bar{\mathbf{L}}(t)=\bar{\mathbf{L}}^T(t)$) or matrix $\bar{\mathbf{L}}(t)$ is balanced (i.e., $\mathbf{1}^T\bar{\mathbf{L}}(t)=\bar{\mathbf{L}}(t) \mathbf{1}\equiv\mathbf{0}$), are assumed, then condition \eqref{eq:boundcondconsis} is trivially satisfied. And we can even get a better result: the estimate sequence is \emph{strongly consistent}. 

  The argument is standard. By Proposition~\ref{pro:propertyL}, we can see that $\bar{\mathbf{L}}(t)$ is a valid Laplacian matrix \cite{OlfFaxMur07}. As a result, $\bar{\mathbf{L}}(t)+\bar{\mathbf{L}}^T(t)$ is positive semidefinite for both cases of the aforementioned symmetric  structures. Following a similar procedure as in the proof of Lemma~5 of \cite{KarMouRam12}, we can prove that $\mathbf{J}_k$ is positive definite in this case, and thus \eqref{eq:boundcondconsis} is satisfied. It then follows from Theorem~\ref{thm:consistency} that the estimate sequence is asymptotically consistent. The strong consistency  can be proved by checking the convergence of the states $V(t_k+l)$, $\forall l=0,1,\dots,\tau-1$, during each interval $[t_k,t_{k+1})$. The proof is quite similar to step 1 of the proof in Appendix~\ref{app:proofconsistency}. So we omit the details. 

\section{Performance-oriented algorithm design for fading channels with channel statistics}\label{sec:algorithmdesign}
  In this section, we will consider the parameter design issue of the proposed algorithm for energy-constrained sensor networks based on the theoretical results established previously. 

  Theorems \ref{thm:asyunb} and \ref{thm:consistency} establish two desired asymptotic properties of the proposed algorithm. The design issue is concerned with choosing appropriate parameters such that the conditions of these theoretical results are satisfied. As \emph{a priori} knowledge, we assume that the channel statistics, i.e., the mean $\bar{h}_{ij}\triangleq\mathbb{E}\{h_{ij}(t)\}>0$, is available at the receiver $i$. 

\subsection{Selection of the decaying weight $a(t)$}
 The first issue in the implementation of the proposed algorithm when applying Theorems~\ref{thm:asyunb} and \ref{thm:consistency} is the selection of weight sequence $\{a(t)\}_{t\geq 0}$. 
 Conditions \eqref{eq:persistence} and \eqref{eq:persistence2} introduced in Theorems~\ref{thm:asyunb} and \ref{thm:consistency} restrict the rate of decrease of $a(t)$. They force the sequence to decrease not too fast and not very slowly. For example, $a(t)=t^{-\alpha}$ with $0<\alpha\leq 1$ satisfies the restriction,  but  does not when $\alpha>1$. This is formally stated in the next result.
 \begin{proposition}\label{pro:weight}
 The function $a(t)=t^{-\alpha}$ with $0<\alpha\leq 1$ satisfies conditions \eqref{eq:persistence} and \eqref{eq:persistence2} of Theorems~\ref{thm:asyunb} and \ref{thm:consistency} with
 $0<\underline{c}\leq (2/3)^{\alpha}$, $0<\kappa<2^{1-\alpha}/(2+\alpha \nu)$ and
\[
  \frac{\kappa\nu t}{t^{1-\alpha}-\alpha\kappa\nu}\leq T(t,\nu)\leq \nu(t-1)^{\alpha}-1,  \ \text{for large}\ t.
\]
  Moreover, if we further require $0.5<\alpha\leq 1$, then condition \eqref{eq:conditionconsis} of Theorem~\ref{thm:consistency} is also satisfied.
 \end{proposition}
\begin{IEEEproof}
See Appendix~\ref{app:proofweight}.
\end{IEEEproof}

\subsection{Selection of parameters $\{\alpha_i(t)\}_{i=1}^N$ and $\{b_i(t)\}_{i=1}^ M$}
  We take two constraints into consideration in designing $\alpha_i(t)$ and $b_i(t)$ of the proposed algorithm \eqref{eq:system} for an energy-constrained network. 
  The first objective is to bring the effect of fading under a satisfactory level, to be concrete, $\mathbb{E}\{\mathbf{\Gamma}(t)\}=\mathbf{0}$ in terms of Theorems \ref{thm:asyunb} and \ref{thm:consistency}. 
  Another aspect is the transmit and received powers, which  should be bounded  at all nodes. Following \cite{LeongDeyEvans11,NoklBajwCaldAazh13}, let us define the transmit power $P_i(t)$ of node $i$ in transmitting its estimate to the neighbors and the received power $R_i(t)$ as 
  \[
  P_{i}(t)\triangleq \alpha_{i}^2(t)\mathbb{E}\{x_i(t)^2\},  \ \text{and} \ R_i(t)\varpropto\sum_{j\in\mathcal{N}_i(t)} P_j(t),
  \]
  respectively.

  The next result gives one choice of such  parameters with desired properties, where only simple scaling rules are adopted.
  \begin{proposition}\label{pro:boundedpower}
   Consider the following scaling rule
 \begin{equation}\label{eq:designalphai}
   \alpha_{i}(t)=\begin{cases}
   d_s^{-0.5}, & i\in\mathcal{I}_s,\\
   d_r^{-0.5}, &i\in\mathcal{I}_r,
   \end{cases}
 \end{equation}
   where $d_s$, $d_r$ are two constants satisfying $d_s\geq \max_{i\in \mathcal{V}} |\mathcal{N}_i^s(t)|$, $d_r\geq \max_{i\in \mathcal{I}_s, j\in \mathcal{I}_r} |\mathcal{N}_i^r(t)||\mathcal{N}_j^s(t)|^2 $, and  the weight $b_{i}(t)=\sum_{j\in\mathcal{N}_i(t)} b_{ij}(t)$, where
\begin{equation}\label{eq:designbij}
   b_{ij}(t)=\begin{cases} d_s^{-0.5}\bar{h}_{ij}, &j\in\mathcal{N}_i^s(t),\\
   (d_sd_r)^{-0.5} \bar{h}_{ij}\sum_{k\in\mathcal{N}_j^s(t)} \bar{h}_{jk}, &j\in\mathcal{N}_i^r(t),\end{cases}
\end{equation}
  then under Assumption 3, we have $\mathbb{E}\{\mathbf{\Gamma}(t)\}=\mathbf{0}$. Moreover, if Assumptions 1, 2 and 4 also hold, and $a(t)$ is set to $a(t)=t^{-\alpha}$ with $0.5<\alpha\leq 1$, then both $P_i(t)$ and $R_i(t)$ are bounded at each node, irrespective of the number of SNs and RNs.
\end{proposition}
\begin{IEEEproof}
See Appendix~\ref{app:proofboundedpower}.
\end{IEEEproof}

 For each node $i\in \mathcal{V}$, it is obvious that $|\mathcal{N}_i^s(t)|\leq M$ and  $|\mathcal{N}_i^r(t)|\leq N-M$, $\forall t\geq 0$. Hence, one possible choice of $d_s$ and $d_r$ in Proposition~\ref{pro:boundedpower} is 
\begin{equation}\label{eq:scalingbound}
  d_s=M,\ \text{and}\  d_r=(N-M)M^2,
\end{equation}
   or any other respective upper bounds of $M$ and $(N-M)M^2$.

\begin{remark}
  One may ask why not use the no scaling rule, i.e., $\alpha_{j}(t) \equiv 1$ for each transmitter $j$. This is the simplest form of the amplification factor. In this case, we can still ensure $\mathbb{E}\{\mathbf{\Gamma}(t)\}=\mathbf{0}$  by designing appropriate $\{b_i(t)\}_{i}$. However, the received power at each node $i$ is $R_i(t)\varpropto \sum_{j\in\mathcal{N}_i^s(t)}\mathbb{E}\{x_i(t)^2\}$, which will grow unbounded as the number of SN neighbors $|\mathcal{N}_i^s(t)|$ goes to infinity. 
\end{remark}

\subsection{Distributed implementation of $\{b_i(t)\}_{i=1}^M$}
 When the channel statistics information  of its two-hop neighbors are not available at SN $i$, the algorithm parameter $b_i(t)$ in \eqref{eq:designbij} is infeasible and has to be modified accordingly. The information $b_{ij}(t)$, $\forall j\in\mathcal{N}_i^s(t)$, for instance is local; but the term $b_{ij}(t)$, $j\in\mathcal{N}_i^r(t)$, depends on the information from two-hop neighbors of SN $i$. 

 One way to tackle the issue arising from two-hop information is to allow RNs to broadcast their available channel statistics  along with the estimates to the neighbors. To be specific, each RN $j$ collects the local channel information $\bar{h}_{j\leftarrow}(t)\triangleq \sum_{k\in\mathcal{N}_j^s(t)} \bar{h}_{jk}$, scales the data using the amplification factor $\alpha_j(t)$ in \eqref{eq:designalphai}, and then broadcast it to its neighbors. Following the multi-access scheme \eqref{eq:multiaccess}, its neighbor SN $i$ receives
\begin{equation*}
b_{i}^r(t)\triangleq \sum_{j\in \mathcal{N}_i^r(t)}\left( d_r^{-0.5} h_{ij}(t)\bar{h}_{j\leftarrow}(t)+v_{ij}(t)\right).
\end{equation*}
 In this case, we can modify the design of $b_{i}(t)$ in Proposition~\ref{pro:boundedpower} as follows
 \begin{equation}\label{eq:designbi2}
  b_i(t)=d_s^{-0.5}\Biggl(\sum_{j\in \mathcal{N}_i^s(t)} \bar{h}_{ij}+b_i^r(t)\Biggr). 
 \end{equation}

 We emphasize that the same conclusions of Proposition~\ref{pro:boundedpower} can be drawn with $\alpha_i(t)$ in \eqref{eq:designalphai} and $b_i(t)$ in \eqref{eq:designbi2}. In fact, by Assumption~3, we have $\mathbb{E}\{v_{ij}(t)\}=0$, $\forall i,j$. This means that 
\[
\mathbb{E}\{b_{i}^r(t)\}=d_r^{-0.5}\sum_{j\in \mathcal{N}_i^r(t)}\bar{h}_{ij}\sum_{k\in\mathcal{N}_j^s(t)} \bar{h}_{jk}.
\]
 As a result, from \eqref{eq:designbi2}, we have $\mathbb{E}\{b_i(t)\}=\mathbb{E}\{\gamma_i(t)\}$, $\forall i$, which implies that $\mathbb{E}\{\mathbf{\Gamma}(t)\}=\mathbf{0}$. Moreover, without much difficulty, we can check that Theorems~\ref{thm:asyunb} and \ref{thm:consistency} hold as well, from which the conclusions of Proposition~\ref{pro:boundedpower} follow.

\begin{figure}[!t]
  \centering
  \includegraphics[width=8.5cm]{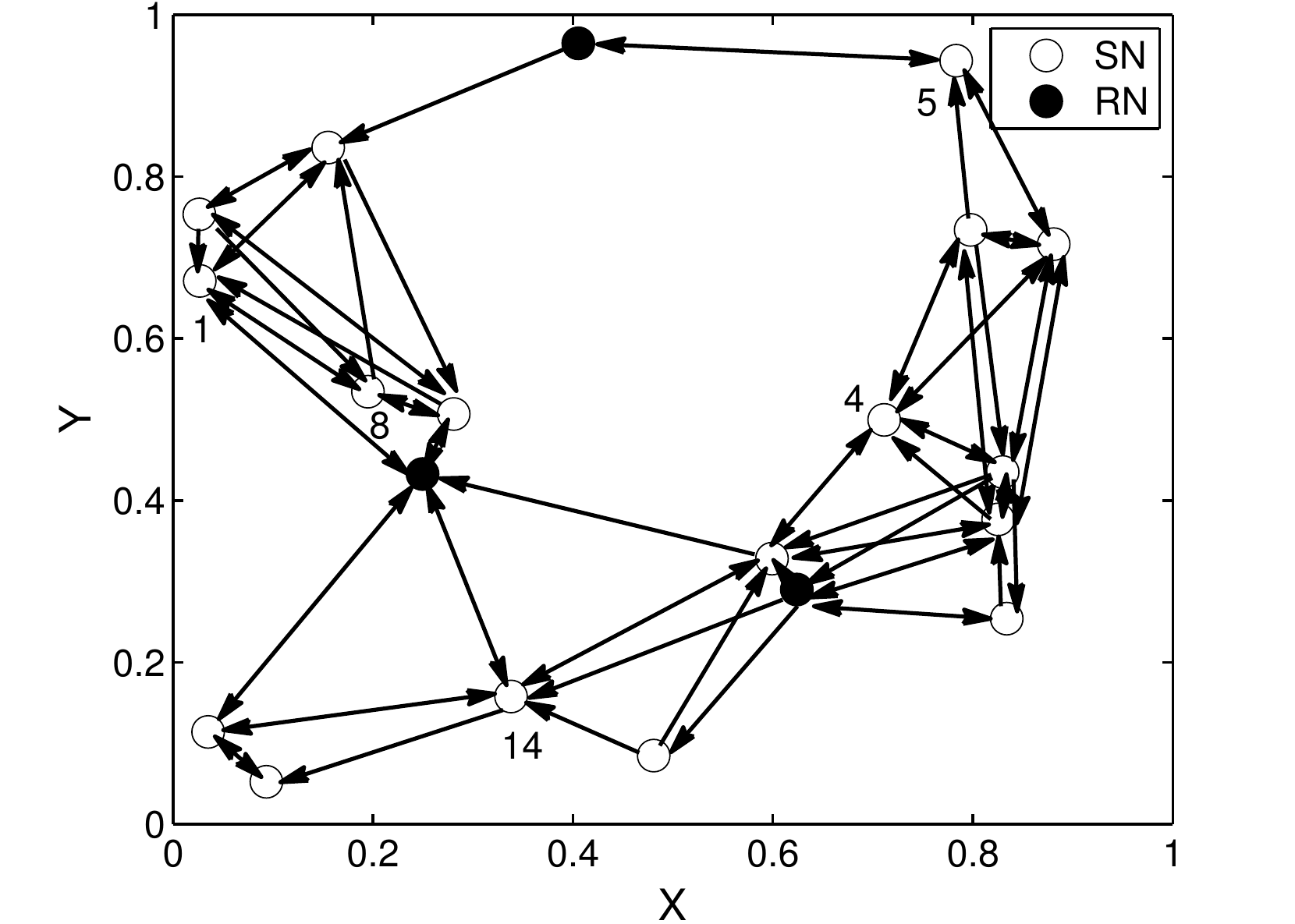}
  \caption{Randomly generated directed network with 20 nodes.}
  \label{fig:directed_graph}
\end{figure}

\section{Numerical studies}\label{sec:simulation}
  In this section, we present some simulation studies of the proposed estimation algorithm to validate the theoretical results and demonstrate the effects of different factors.

  We consider a network of $N=20$ nodes with $M=17$ SNs and 3 RNs to monitor an unknown $\theta=2$. The network is obtained using the random geometric graph model, i.e., nodes are placed uniformly at random over the unit square $[0,1]\times [0,1]$. We connect two nodes by a link if their distance is less than $\sqrt{\log N/N}$. After the initial deployment, we randomly remove 30\% of the unidirectional links to generate a directed graph, whose ideal topology is shown in Fig.~\ref{fig:directed_graph}. To simulate the time-varying feature of the topology, we let 10\% of links of the ideal topology fail at every time instant during the simulations.

  We assume that only 20\% of SNs can measure $\theta$ every $T=5$ steps with zero-mean noise $w_i(t)$ whose variance $\sigma_{w}^2=0.5$.
  We model the fading effect of communications as i.i.d. Rayleigh distributions with the parameter $\sigma_{ray}=1$. The receiver noises are  i.i.d. Gaussian with variance $\sigma^2_{v}=0.02$. We choose $\alpha_i(t)$ and $b_{ij}(t)$  according to Proposition~\ref{pro:boundedpower} and \eqref{eq:scalingbound}, and $a(t)=(t+1)^{-0.7}$.
  Moreover, the  initial estimates $x_{i}(0)$ of SNs are randomly chosen from the interval $[-4,8]$, but remain fixed for all runs. In the following, simulation results are presented by averaging over 100 independent runs.

\begin{figure}[!t]
  \centering
  \subfloat[]{\includegraphics[width=8.5cm]{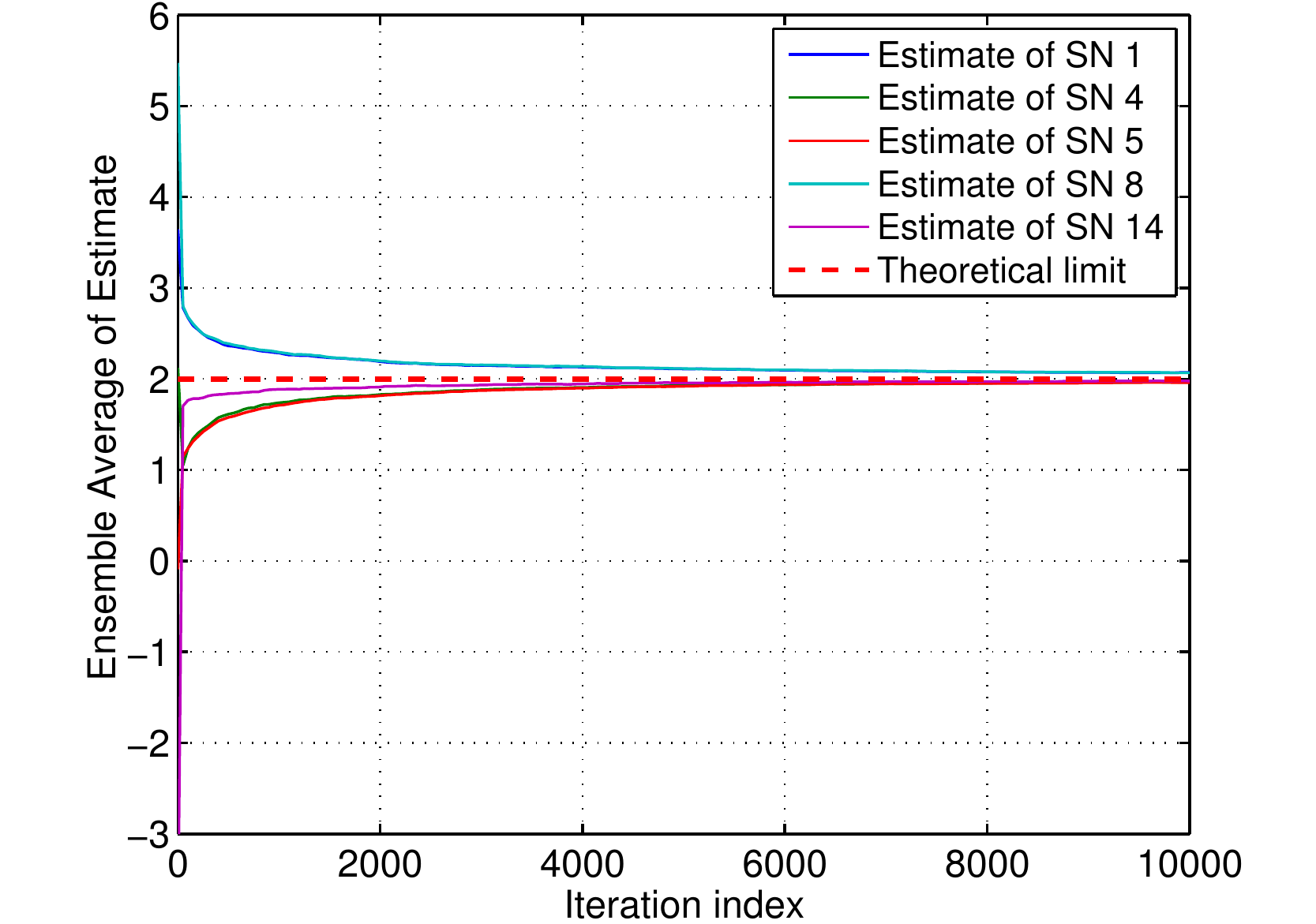}}
  \subfloat[]{\includegraphics[width=8.5cm]{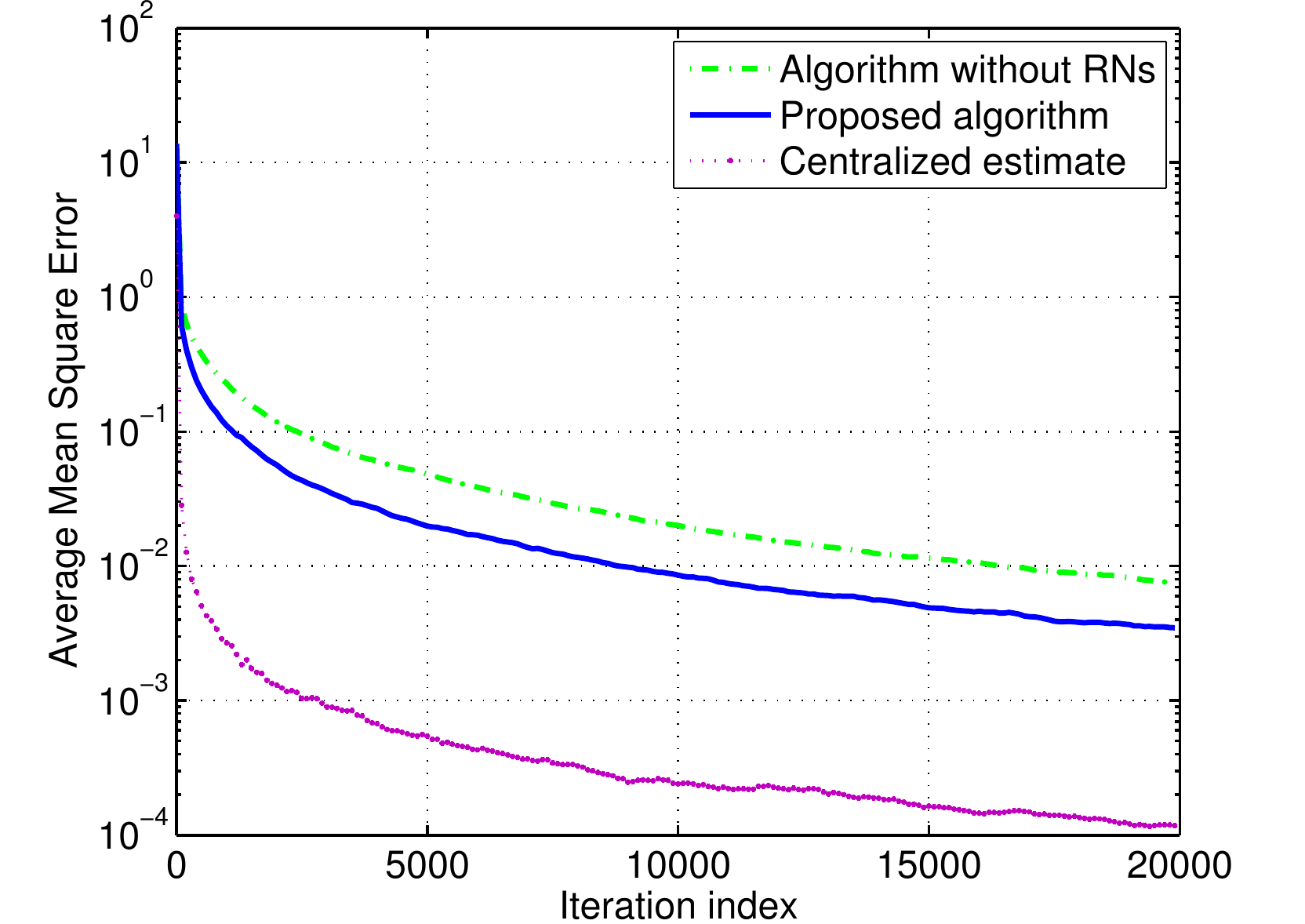}}
  \caption{(a) Ensemble average of the estimate $x_i(t)$ for node $i=1,4,5,8,14$; (b) Comparison of average mean square error $e_{\text{mse}}(t)$ of the proposed algorithm with/without RNs and the centralized estimator.}
  \label{fig:ensembleaverage}
\end{figure}

  Fig.~\ref{fig:ensembleaverage}(a) depicts the ensemble average of the estimate $x_i(t)$ for nodes $i=1,4,5,8,14$ for  illustration purpose. Also plotted is the theoretical limit of the proposed algorithm, i.e., the true value $\theta=2$. As we can notice from the figure, each ensemble average closely approaches the theoretical value, which corroborates the theoretical result  obtained in Theorem~\ref{thm:asyunb}. Moreover, nodes within the same group, i.e., $\{4,5\}$ and $\{1,8\}$, achieve an agreement on the estimate faster. This is because nodes in the same group have stronger coupling strength than those in different groups. In Fig.~\ref{fig:ensembleaverage}(b), we illustrate the performance of the algorithm via the average mean square error $e_{\text{mse}}(t)\triangleq (1/M)\|\mathbf{x}(t)-\theta \mathbf{1}\|_2^2$.
  We also plot the results of the proposed algorithm without RNs and the centralized estimate for the purpose of comparison. For the centralized estimate, we assume that a fusion center has access to all the measurements, the amplification factors, and the channel statistics. We remark that the proposed algorithm without RNs reduces to the one proposed in \cite{KarMouRam12} in the absence of channel fading. From  Fig.~\ref{fig:ensembleaverage}(b), we can see that $e_{\text{mse}}(t)$ goes to zero as $t\to \infty$, thus conforming the theoretical result obtained in Theorem~\ref{thm:consistency}. We can also see that the introduction of a few number of RNs in the network would result in improved performance compared with the one without RNs. This is clear by taking into consideration the fact that the directed network without RNs is disconnected (see the ideal network shown in Fig.~\ref{fig:directed_graph}).
  The result indicates that inserting several RNs in the network would be beneficial to the performance of the algorithm. 

\begin{figure}[!t]
  \centering
  \subfloat[]{\includegraphics[width=8.5cm]{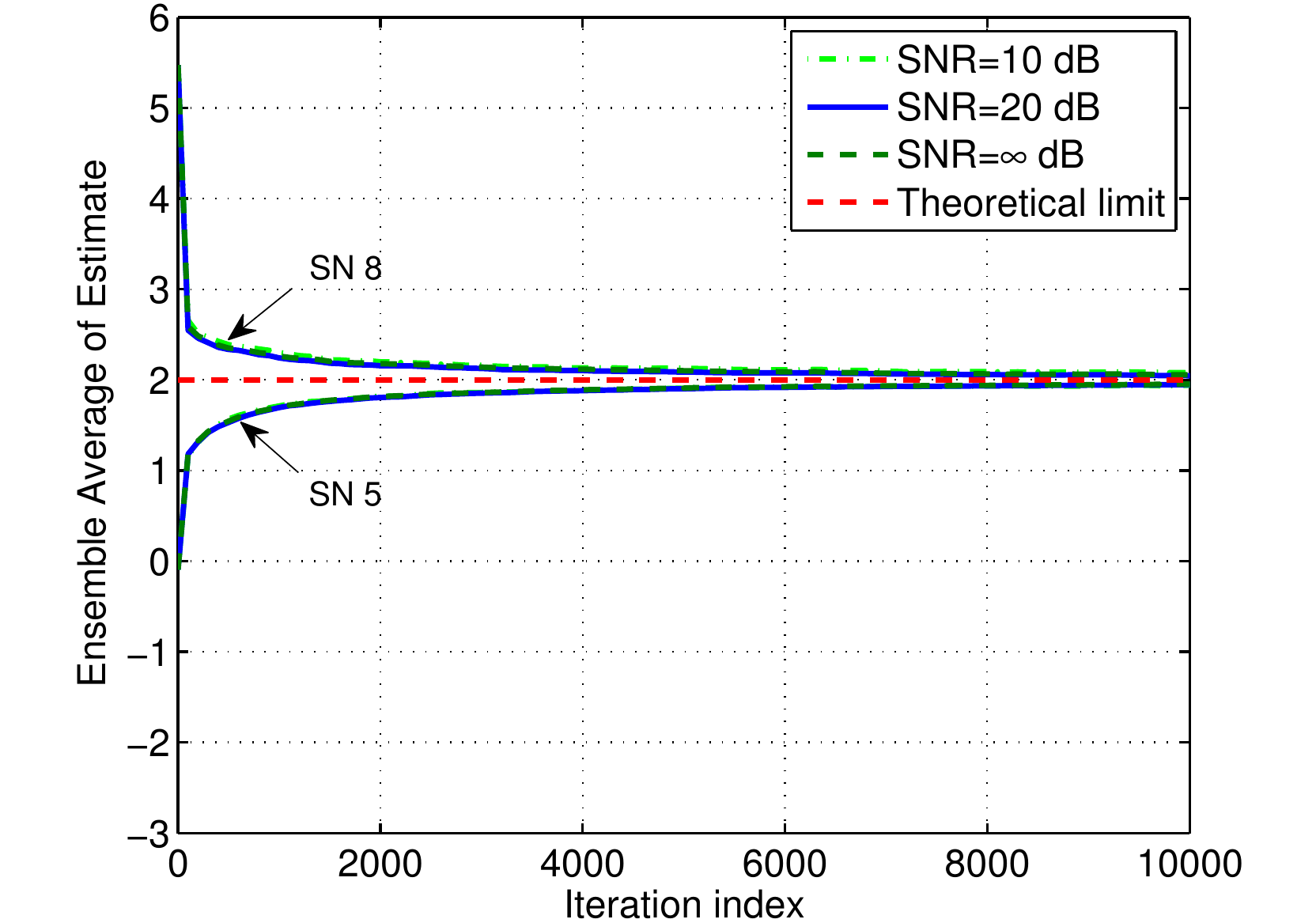}}
  \subfloat[]{\includegraphics[width=8.5cm]{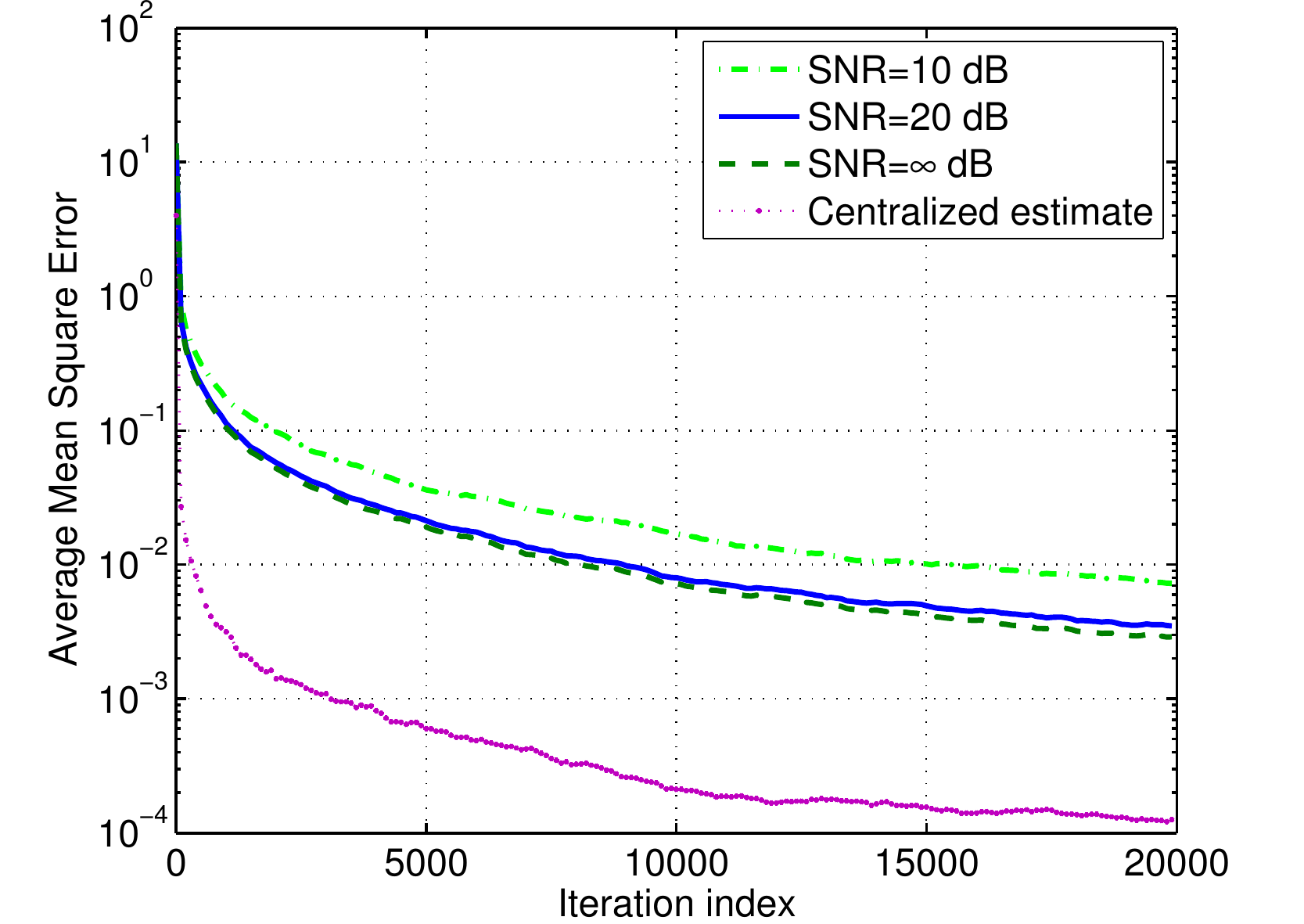}}
  \caption{(a) Ensemble average of the estimate $x_i(t)$, $i=5,8$, and (b)  average mean square error $e_{\text{mse}}(t)$ for different SNRs with SNR=10, 20 and $\infty$ dB.}
  \label{fig:xMSE_SNR}
\end{figure}

 In Fig.~\ref{fig:xMSE_SNR}, we plot the ensemble average of the estimate $x_i(t)$, $i=5, 8$, and $e_{\text{mse}}(t)$ for different reception signal-to-noise ratios (SNRs), which is defined as $\text{SNR}=10\log_{10}(\theta/\sigma_v^2)$. We adjust the receiver noise variance $\sigma_v^2$ so that SNR=10~dB, and 20 dB. As a comparison, we also provide the case that SNR=$\infty$ dB, i.e., no receiver noises. Clearly, the proposed algorithm solves the estimation problem for different SNRs, achieving both the asymptotic unbiasedness and consistency. The results indicate that the proposed algorithm exhibits resilience to communication noise. It is also observed that as the SNR increases, the estimation error reduces. Whereas for higher SNRs, there are only marginal improvements. This is mainly due to the effect of channel fading, which has a great impact on the estimation performance of the algorithm even if there no receiver noises.

\begin{figure}[!t]
  \centering
  \subfloat[]{\includegraphics[width=8.5cm]{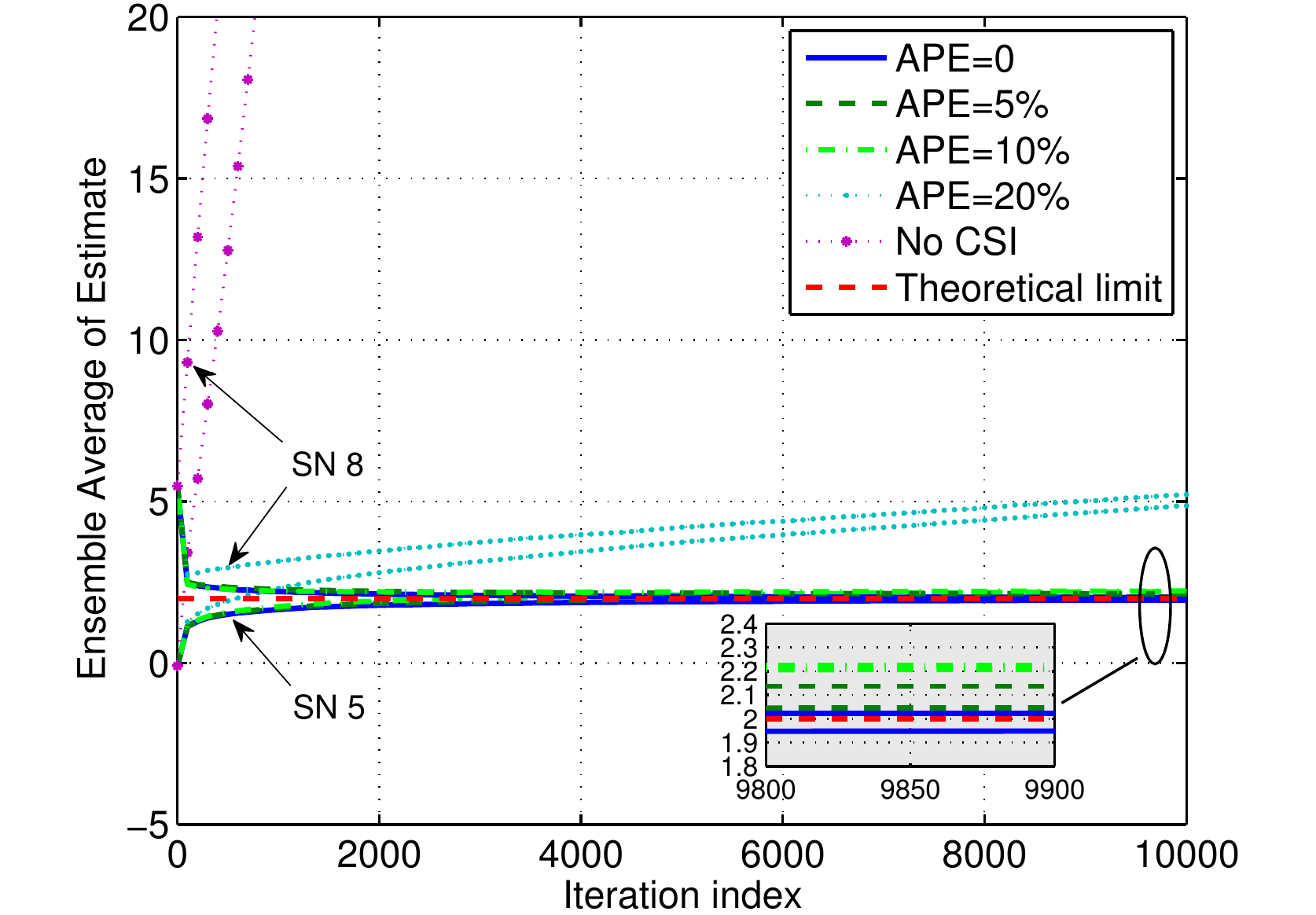}}
  \subfloat[]{\includegraphics[width=8.5cm]{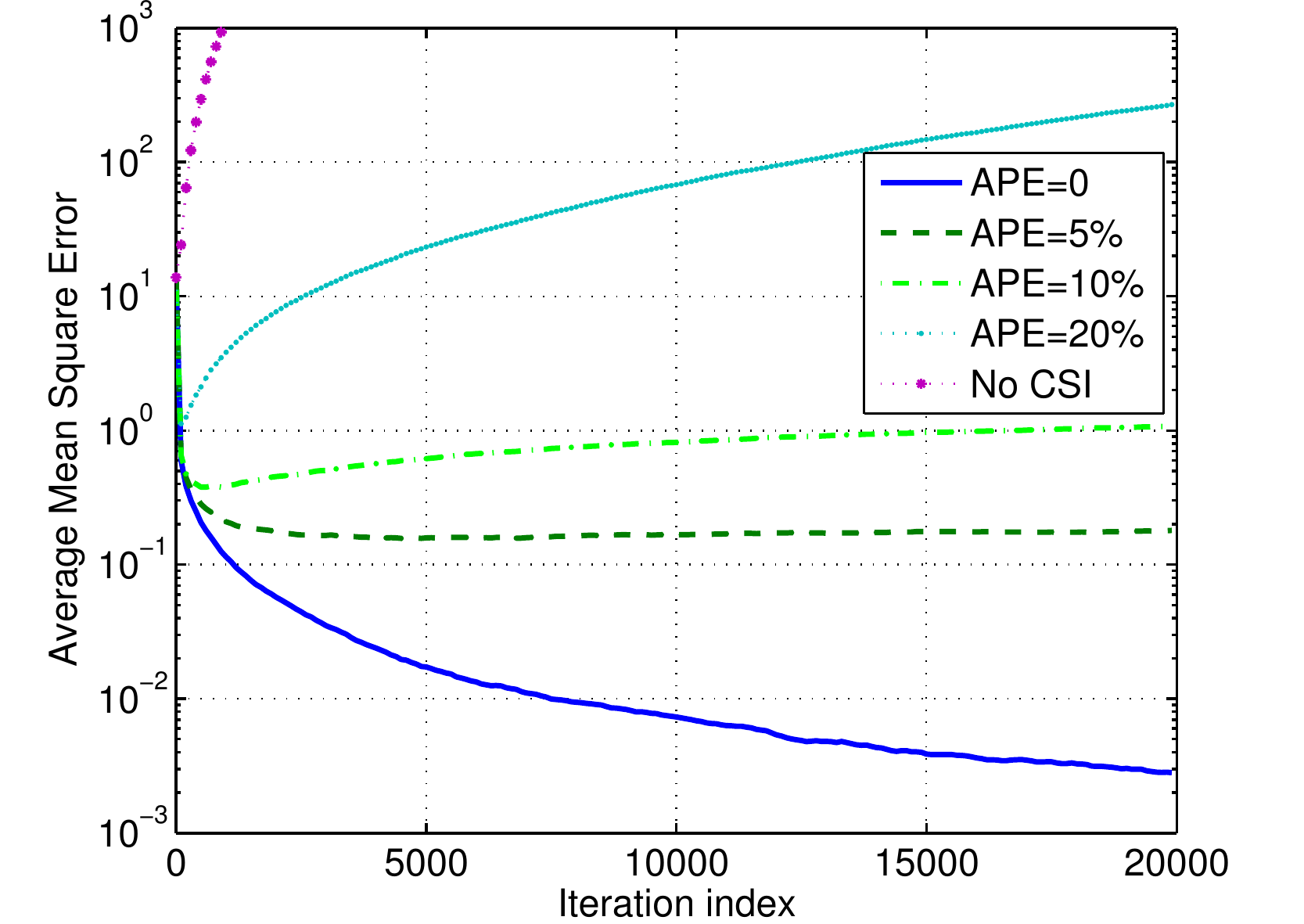}}
  \caption{(a) Ensemble average of the estimate $x_i(t)$, $i=5,8$, and (b)  average mean square error $e_{\text{mse}}(t)$ for different levels of knowledge of the channel statistics.}
  \label{fig:xMSE_fading}
\end{figure}

  In Fig.~\ref{fig:xMSE_fading}, we test the performance of the algorithm for different levels of knowledge of channel statistics. We model the uncertainty of the channel statistics $\bar{h}_{ij}$ at each node by percentage error (PE). In each run of simulation, $\bar{h}_{ij}$ is randomly generated, which is corrupted by uniform noise. We take the average of all PEs, and get $\text{APE}=1/(\#\text{runs}) \sum_{k=1}^{\# \text{runs}} \text{PE}(k)$. If no \emph{a priori} knowledge of the channel statistics information (CSI) is assumed at the nodes, we use $\bar{h}_{ij}=1$ in $b_i(t)$ of \eqref{eq:designbij}. This is the procedure for distributed estimation in the absence of channel fading \cite{KarMouRam12,StanSti11,ZhangZhang12,ZhuChenMaYangGuan15}. The simulation results in Fig.~\ref{fig:xMSE_fading} demonstrate that the proposed algorithm performs well for smaller APEs. In particular, the estimates for both APE=5\% and 10\% are satisfactory with respect to the first moment (see Fig.~\ref{fig:xMSE_fading}(a)). As for $e_{\text{mse}}(t)$, it  tends to be bounded for APE=10\%. However, when APE increases to 20\%,  $e_{\text{mse}}(t)$ is quite large and unacceptable. These results indicate that although the proposed algorithm relies on the exact knowledge of CSI as in Proposition~\ref{pro:boundedpower}, it is resilient to CSI uncertainty to some extent. 
  We also notice that both the ensemble average of the estimate  and $e_{\text{mse}}(t)$ grow unbounded in a few iterations, if no \emph{a priori} knowledge of CSI is assumed. This shows that the schemes  in \cite{KarMouRam12,StanSti11,ZhangZhang12,ZhuChenMaYangGuan15} are no longer applicable in the presence of channel fading, and much attention should be paid to the algorithm design for estimation problems in networks affected by channel fading.

\begin{figure*}[!t]
  \centering
  \includegraphics[width=8.5cm]{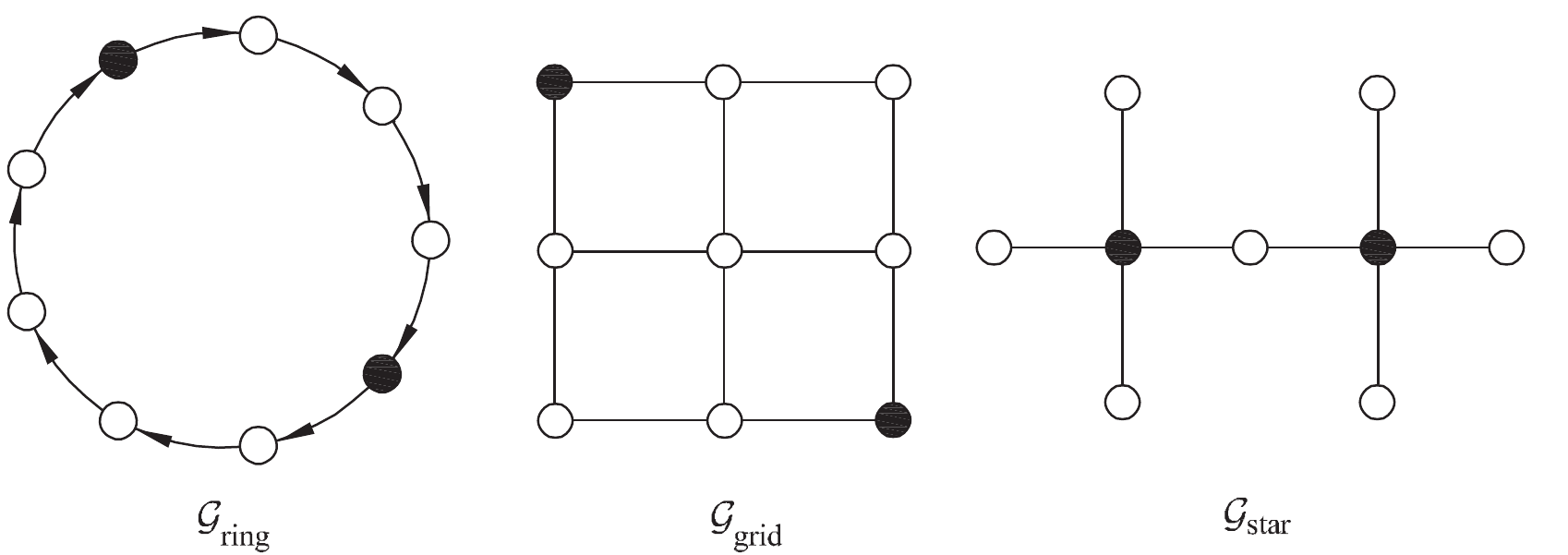}
  \caption{Network topologies used for comparison.}
  \label{fig:topology}
\end{figure*}

\begin{figure*}[!t]
  \centering
  \subfloat[]{\includegraphics[width=8.5cm]{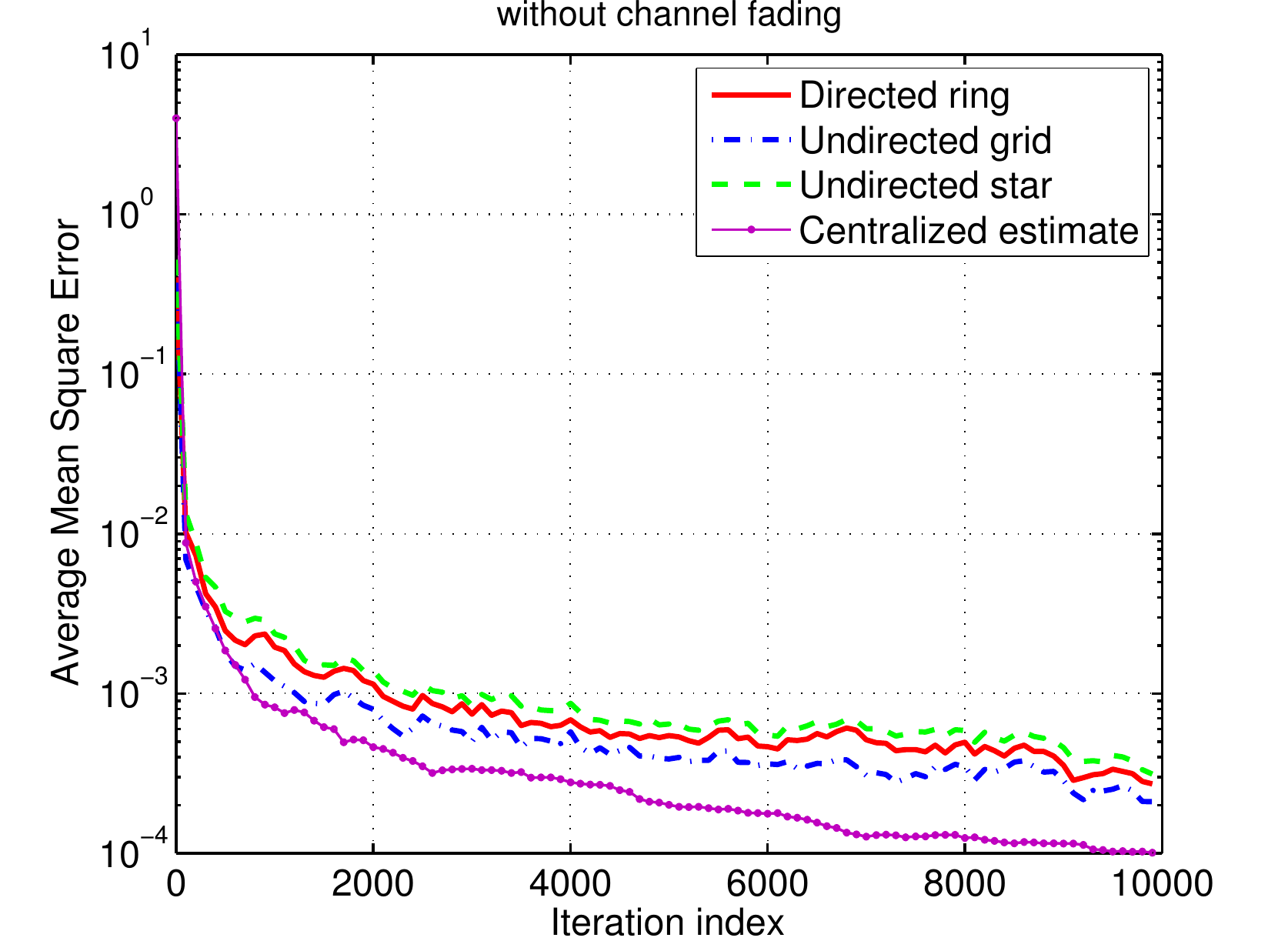}}
  \subfloat[]{\includegraphics[width=8.5cm]{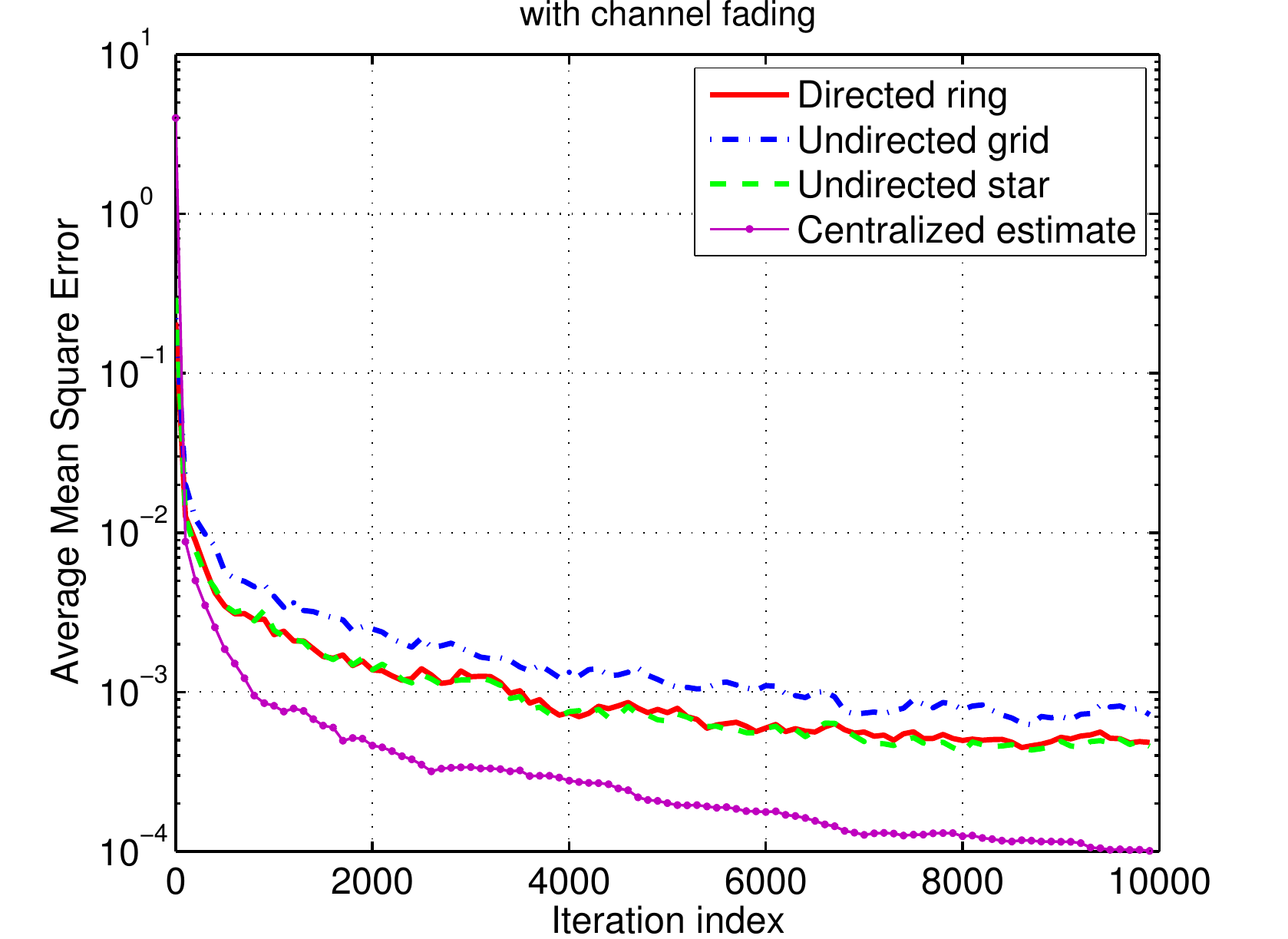}}\\
  \subfloat[]{\includegraphics[width=8.5cm]{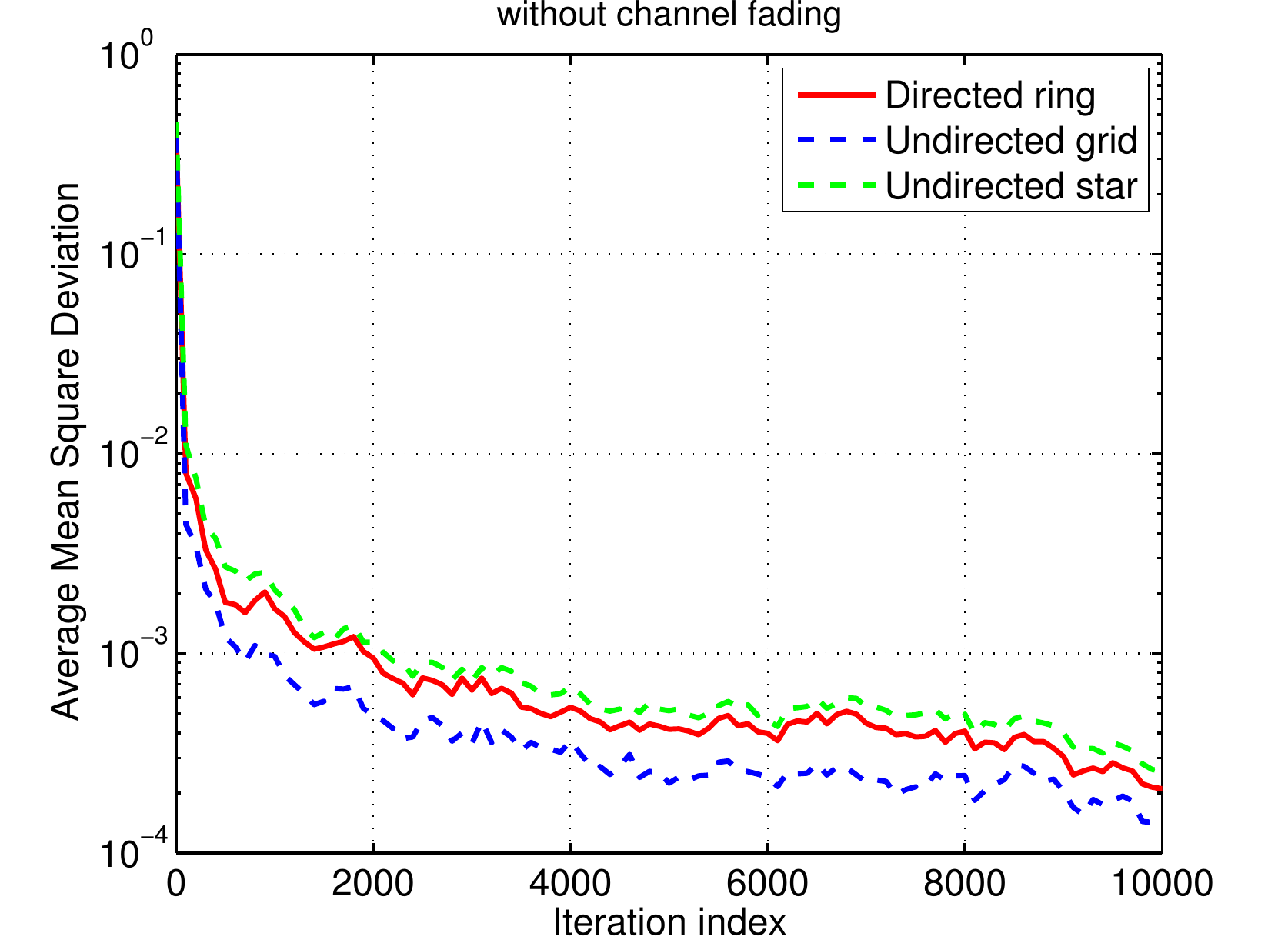}}
  \subfloat[]{\includegraphics[width=8.5cm]{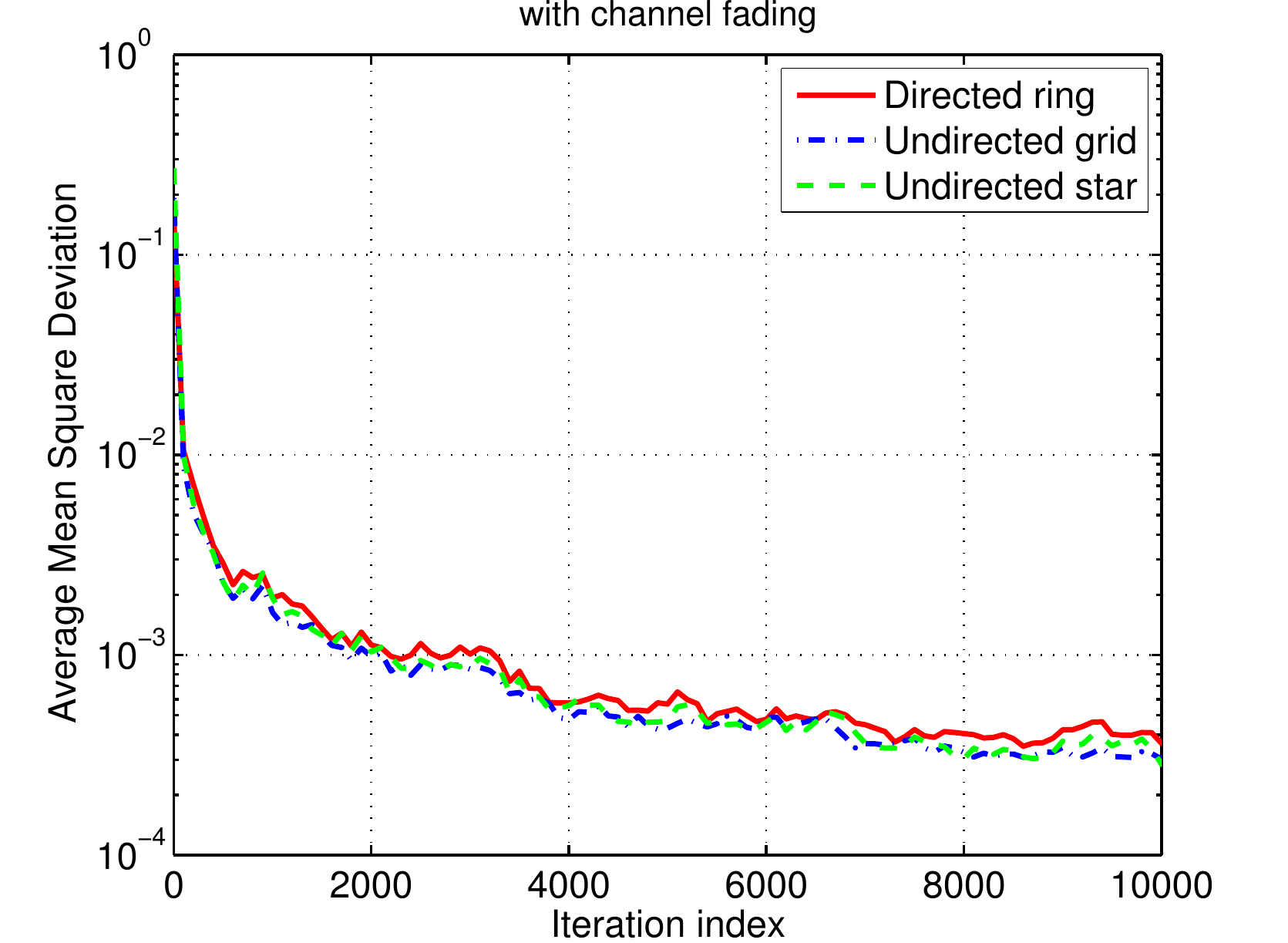}}
  \caption{Comparison of three different topologies without/with channel fading: average mean square error $e_{\text{mse}}(t)$ (top two figures) and average mean square deviation $e_{\text{msd}}(t)$ (bottom two figures).}
  \label{fig:comparison_topo}
\end{figure*}

  Next, let us examine the effect of network topology and channel fading.
  We consider three different communication topologies composed of $M=7$ SNs  and 2 RNs, whose ideal topologies are shown in Fig.~\ref{fig:topology}. For the simulation,  there are two links that fail randomly at each step for each topology. And there is one SN that can measure the unknown parameter every $T'=3$ time steps, and a second SN that takes measurement every $T''=4$ time steps. In Fig.~\ref{fig:comparison_topo}(a) and (b), we plot $e_{\text{mse}}(t)$ for the three different topologies when there are/aren't channel fading. It can be observed that in the absence of channel fading, $\mathcal{G}_{\text{grid}}(t)$ gives the smallest $e_{\text{mse}}(t)$, followed by $\mathcal{G}_{\text{ring}}(t)$ and $\mathcal{G}_{\text{star}}(t)$. The reason is that  $\mathcal{G}_{\text{grid}}$ has larger graph density than $\mathcal{G}_{\text{ring}}$ and $\mathcal{G}_{\text{star}}$. 
  Actually, $\mathcal{G}_{\text{grid}}$ has 10 links, while $\mathcal{G}_{\text{ring}}$ and $\mathcal{G}_{\text{star}}$ have 9 and 8 links, respectively. 
  This  means that  $\mathcal{G}_{\text{grid}}(t)$ has the strongest coupling strength between nodes. And we know that stronger coupling enables fast propagation of information over the  network. However, when the communication between nodes suffers from channel fading, completely different behaviors are noticed in Fig.~\ref{fig:comparison_topo}(b), where stronger coupling has an adverse effect on the estimation accuracy.


  At first sight, this result seems strange. Actually, it is reasonable if we take the following facts into account:
  (i) Our algorithm~\eqref{eq:stamodeI}-\eqref{eq:stamodeII} assumes that the measurement rate and communication rate are similar. 
  For a long time period, even if the coupling between nodes is weak, all the nodes in the network can still obtain sufficiently rich information from the environment as long as the network is strongly connected;
  (ii) Communications between nodes are exposed to channel fading and noise. These undesired effects have a great impact on the convergence of the algorithm. Stronger coupling means more noises are incurred in the algorithm.
  Hence, the relative gain of each transmission between nodes will be soon outperformed by the loss of performance caused by the adverse effects. 

  Another measure of the performance of the algorithm is the disagreement of estimates of the nodes \cite{Olf07}. We use the average mean square deviation  $e_{\text{msd}}(t)\triangleq (1/M) \|\mathbf{x}(t)-x_{\text{ave}}\mathbf{1}\|_2^2$ as the indicator, where $x_{\text{ave}}=(1/M)\sum_{i=1}^M x_i(t)$.  The simulation results of $e_{\text{msd}}(t)$ without/with channel fading are shown in Fig.~\ref{fig:comparison_topo}(c) and Fig.~\ref{fig:comparison_topo}(d), respectively. In both cases, $\mathcal{G}_{\text{grid}}(t)$ produces the smallest $e_{\text{msd}}(t)$, which is consistent with the intuition that stronger coupling between nodes gives smaller disagreement of estimates among the nodes. To sum up, the results in Fig.~\ref{fig:comparison_topo} reveal that we should pay much attention to balance the performances  $e_{\text{mse}}(t)$ and $e_{\text{msd}}(t)$ in networks with fading channels by deploying appropriate topologies.

\begin{figure}[!t]
  \centering
  \includegraphics[width=8.5cm]{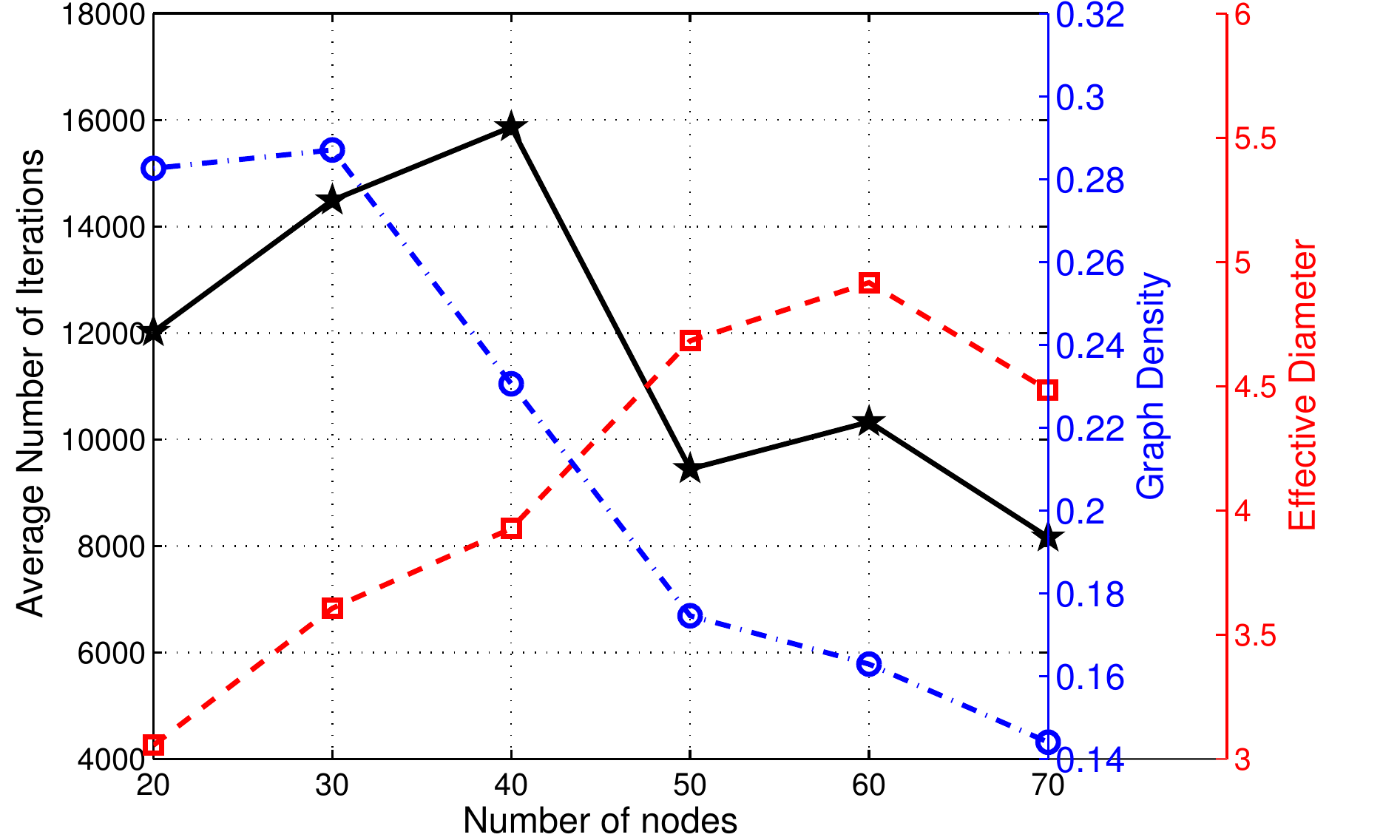}
  \caption{Average number of iterations to get to $e_{\text{mse}}(t)$ of $10^{-2}$  along with the graph density and effective diameter versus the number of nodes.}
  \label{fig:no_iteration}
\end{figure}

  Finally, we investigate the average number of iterations against the number of nodes in the network. We randomly generate several undirected network topologies using the random geometric graph model, and then designate 15\% of the nodes as RNs. Other settings remain the same as in the first simulation. Fig.~\ref{fig:no_iteration} depicts the average number of iterations to get to $e_{\text{mse}}(t)$ of $10^{-2}$ when the number of nodes in the network varies from $N=20$ to 70. It is noticed that $e_{\text{mse}}(t)$ does not simply increase with $N$ in the presence of channel fading. In order to find out the best correlated quantities of the network topology with such behavior, we also plot the graph densities and effective diameters of the corresponding ideal topologies in  Fig.~\ref{fig:no_iteration}. Here the graph density is defined as $number\ of\ links/[M(N-(M+1)/2)]$, where the denominator is the maximum number of links in the relay-assisted network topology. And the effective diameter is the minimum value $d$ such that at least 90\% of the connected node pairs are at distance at most $d$ \cite{LesKleFal05}. What Fig.~\ref{fig:no_iteration} indicates is that the effective diameter has a great impact on the time to convergence, which follows a similar behavior as that of the number of iterations within some range. However, when the effective diameter increases to some extent (around 4), the number of iterations will drop to a low level. From Fig.~\ref{fig:no_iteration}, we can identify that this transition is greatly related with the graph density as can be observed from Fig.~\ref{fig:no_iteration}. 
  This result is consistent with those reported in Fig.~\ref{fig:comparison_topo}, where graph density is also a crucial factor for the performance of the proposed algorithm over different topologies.

\section{Conclusions}\label{sec:conclusion}
  A general problem of distributed inference in relay-assisted sensor networks was considered. Some practical issues such as time-varying asymmetric topology, intermittent arrival of new measurements, and channel fading were taken into account. We proposed a distributed estimation algorithm based on innovation schemes.  We first established the general results regarding asymptotic unbiasedness and consistency of the algorithm by resorting to the ordering technique and the generalized Perron complement. Then we presented the performance-oriented algorithm design in an energy-constrained network based on the theoretical results. Finally, simulation results were given to validate the effectiveness of the proposed algorithm. We also investigated the effects of network topology, channel fading and some other topology-relevant quantities on the performance of the proposed algorithm.

\appendices
\section{Proof of Proposition~\ref{pro:propertyL} } 
\label{app:proofpropertyL}
 i) Under Assumptions 2-4, $\mathbf{w}(t)$, $\mathbf{v}(t)$, $\mathbf{L}(t)$ and $\mathbf{\Gamma}(t)$ are independent of $\mathcal{F}_t$. Thus $\{\mathbf{x}(t), \mathcal{F}_t\}_{t\geq 0}$ is a Markov process.

 ii) The fact that $\mathbf{L}(t)\mathbf{1}=\mathbf{0}$ is obvious by the definition of $\mathbf{L}(t)$. Moreover, we note that, under Assumption~3,  $\alpha_k(t) \bar{h}_{ik} \bar{h}_{kj}\geq 0, \ \forall k\neq i,j$. 
  This implies that $\mathbb{E}\{l_{ij}(t)\}\leq 0$, $\forall j\neq i$. Actually, we can further obtain 
 \[
 \mathbb{E}\{l_{ij}(t)\}\leq
 \begin{cases}
 -\alpha_j(t)\bar{h}_{ij}, &(j,i)\in \mathcal{E}(t),\\
 -\alpha_j(t)\alpha_p(t)\bar{h}_{ip}\bar{h}_{pj}, & (j,p), (p,i)\in \mathcal{E}(t).
 \end{cases}
 \]
   This completes the proof.

\section{Proof of Proposition~\ref{pro:irredu}}\label{app:proofirredu}
 By expanding, we have
\begin{equation}\label{eq:transitionmatrix}
  \mathbf{\Psi}_{t_{k+1},t_k}=\mathbf{I}-\sum_{s=t_k}^{t_{k+1}-1}a(s)\mathbf{\Phi}(s)+ \sum_{s_1>s_2}a(s_1)a(s_2)\mathbf{\Phi}(s_1)\mathbf{\Phi}(s_2)
  -\dots+(-1)^\tau \prod_{s=t_k}^{t_{k+1}-1} a(s)\mathbf{\Phi}(s).
\end{equation}
  Since $a(t)$ can be sufficiently small and $\mathbb{E}\{\mathbf{\Gamma}(t)\}=\mathbf{0}$, the dominant term of $\mathbb{E}\{\mathbf{\Psi}_{t_{k+1},t_k}\}$ is $\mathbf{I}-\sum_{s=t_k}^{t_{k+1}-1}a(s)\mathbb{E}\{\mathbf{\Phi}(s)\}=\mathbf{I}-\sum_{s=t_k}^{t_{k+1}-1}a(s)(\mathbf{\Delta}(s)+\mathbb{E}\{\mathbf{L}(s)\})$.

  In view of Proposition~\ref{pro:propertyL}, we know that $\mathbb{E}\{l_{ij}(t)\}\leq 0$, and  $\mathbb{E}\{l_{ij}(t)\}<0$ if and only if node $j$ can talk to node $i$ directly or via one-hop RNs. By Assumption~1, 
  $\mathcal{G}_{[t_k,t_{k+1})}$ is strongly connected on average. 
  Hence,  $\sum_{s=t_k}^{t_{k+1}-1}\mathbb{E}\{\mathbf{L}(s)\}$ is irreducible \cite[p.78]{Minc88}, which implies that  $\mathbf{I}-\sum_{s=t_k}^{t_{k+1}-1}a(s)(\mathbf{\Delta}(s)+\mathbb{E}\{\mathbf{L}(s)\})$  is irreducible as well. This completes the proof. 


\section{Proof of Lemma~\ref{lem:propertiesofz}}\label{app:proofpropertiesofz}
  Since $\mathbb{E}\{\mathbf{\Gamma}(t)\}=\mathbf{0}$, Eq. \eqref{eq:1stmoment} reduces to 
\begin{equation}\label{eq:errordynCSI}
 \bar{\mathbf{e}}(t+1)=\bigl[\mathbf{I}-a(t)(\mathbf{\Delta}(t)+\bar{\mathbf{L}}(t))\bigr]\bar{\mathbf{e}}(t).
\end{equation}
By \eqref{eq:condition_at}, we have  $a(t)(\chi_i(t)+\bar{l}_{ii}(t))\leq 1$, $\forall i, t\geq t^*$. This together with Proposition~\ref{pro:propertyL} yields $|1-a(t)(\chi_i(t)+\bar{l}_{ii}(t))|+\sum_{j\neq i}|\bar{l}_{ij}(t)|=1-a(t)\chi_i(t)$, $\forall i, t\geq t^*$. Hence, one obtains
\[
   \|\mathbf{I}-a(t)(\mathbf{\Delta}(t)+\bar{\mathbf{L}}(t))\|_\infty=\max_{i}(1-a(t)\chi_i(t))\leq 1, \forall t\geq t^*,
\]
   which gives $\|\bar{\mathbf{e}}(t+1)\|_\infty\leq  \|\bar{\mathbf{e}}(t)\|_\infty$, $\forall t\geq t^*$.
  It thus follows  that $\max_{i}|z_i(t)|\leq \|\bar{\mathbf{e}}(t^*)\|_\infty$.
  Let $c_0\triangleq \max_{0\leq t\leq t^*} \|\bar{\mathbf{e}}(t)\|_\infty$, then i) follows.

   Now we turn to the proof of ii). It follows from \eqref{eq:errordynCSI} and Proposition~\ref{pro:propertyL} that 
\begin{align*}
   \bar{e}_i(t+1)
    &=[1-a(t)(\chi_i(t)+\bar{l}_{ii}(t))] \bar{e}_i(t)-a(t)\!\sum_{j\neq i} \bar{l}_{ij}(t)\bar{e}_j(t)\\
   &\overset{(a)}{\leq} [1-a(t)(\chi_i(t)+\bar{l}_{ii}(t))]z_1(t)-a(t)\!\sum_{j\neq i} \bar{l}_{ij}(t)z_1(t)\\
   &\overset{(b)}{=}(1-a(t)\chi_i(t)) z_1(t),\  \forall t\geq t^*,
\end{align*}
 where $(a)$ follows from \eqref{eq:condition_at} and the fact that $\bar{l}_{ij}\leq 0$, $\forall j\neq i$, and $(b)$ is obtained using the property $\sum_{j=1}^M \bar{l}_{ij}(t)=0$, $\forall i$. Similarly, we can obtain the lower bound of $\bar{e}_i(t+1)$. Hence, for all $t\geq t^*$, one has for each $i$,
 \begin{equation}\label{eq:boundMeanE}
  (1-a(t)\chi_i(t))z_M(t)\leq \bar{e}_i(t+1)\leq (1-a(t)\chi_i(t)) z_1(t).
 \end{equation}
    We consider the following three cases. First, it is noted from  \eqref{eq:condition_at} and Proposition~\ref{pro:propertyL} that $a(t)\leq 1$, $\forall t\geq t^*$.

    \emph{Case 1)}. $z_1(t^*)\leq 0$.   Continuing the right inequality of \eqref{eq:boundMeanE}, we have 
    \[
    z_1(t)\leq \prod_{s=t^*}^{t-1}(1-a(s)\chi_{1_{s+1}}(s)) z_1(t^*), \ \ \forall t>t^*.
    \]
    Hence $z_M(t)\leq z_1(t)\leq 0$, $\forall t\geq t^*$. This together with the left inequality of \eqref{eq:boundMeanE} yields  $\bar{e}_i(t+1)\geq z_M(t)$,  $\forall i$, and thus $z_M(t+1)\geq z_M(t)$, $\forall t\geq t^*$, which means that $z_M(t)$ is  nondecreasing in $t$ on the interval $[t^*,\infty)$. Hence,  there exists a finite $z_{M\infty}\leq 0$ such that $\lim_{t\to \infty}z_M(t)=z_{M\infty}$.

     \emph{Case 2)}. $z_M(t^*)\geq 0$. In this case, using the left inequality of \eqref{eq:boundMeanE}, one has 
     \[
     z_M(t)\geq \prod_{s=t^*}^{t-1} (1-a(s)\chi_{M_{s+1}}(s))z_M(t^*),
     \] 
    and then $z_1(t)\geq z_M(t)\geq 0$, $\forall t\geq t^*$.
    Thus from the right inequality of \eqref{eq:boundMeanE}, we have $\bar{e}_i(t+1)\leq z_1(t)$, $\forall i$, which shows that $z_1(t+1)\leq z_1(t)$, $\forall t\geq t^*$. Consequently, there exists a finite $z_{1\infty}\geq 0$ such that $\lim_{t\to \infty}z_1(t)= z_{1\infty}$.

     \emph{Case 3)}. $z_M(t^*)<0<z_1(t^*)$. We have either $z_1(t^{**})\leq 0$ for some time $t^{**}>0$ or $z_1(t)\geq 0$ for all $t\geq t^*$. For the former case,  it is reduced to Case 1) on the interval $[t^{**}, \infty)$. As for the latter case, the proof is similar to that of  Case 2).

      Combining the above three cases, we complete the proof.

\section{Proof of Theorem~\ref{thm:asyunb}}\label{app:proofunbiasedness}
   We divide the proof into two steps:

  \textbf{Step 1:} (Consensus of $\{\bar{e}_i(t)\}_{1\leq i\leq M}$)
  Consider the recursion of $\bar{\mathbf{e}}(t)$ given in \eqref{eq:errordynCSI}, we have the following claim. 

\emph{Claim:} All the states $\bar{e}_i(t), 1\leq i\leq M$, converge to a common finite limit $z_\infty$, i.e., $\lim_{t\to\infty} \bar{\mathbf{e}}(t)=z_{\infty} \mathbf{1}$.

The proof is inspired by some ideas from \cite{Huang12}. The basic idea is to use the joint connectivity condition to show that all states $z_i(t)$, $1\leq i\leq M$, converge to the same value, given the convergence of $z_1(t)$ or $z_M(t)$ in Lemma~\ref{lem:propertiesofz}. The proof is rather technical and for this reason is postponed to Appendix~\ref{app:proofclaim}.

 \textbf{Step 2:} (Achieving unbiasedness) We show that $z_{\infty}=0$.
We use a contradiction argument. Suppose that $z_{\infty}\neq 0$, then by the Claim in step 1, for sufficiently small $\epsilon>0$, we can find a constant $t_{*}>0$ such that for each $1\leq i\leq M$, 
\begin{equation}\label{eq:boundbare}
 0<|z_{\infty}|-\epsilon\leq |\bar{e}_i(t)|\leq |z_{\infty}|+\epsilon, \ \forall t\geq t_{*}.
\end{equation}
Since $\lim_{t\ \to\infty}a(t)=0$  by \eqref{eq:persistence}, one has
\begin{equation}\label{eq:boundrho0}
 a(t)\leq 1<\mu\triangleq \frac{M(|z_{\infty}|+\epsilon)}{|z_{\infty}|-\epsilon}, \ \forall t\geq t_{*}.
\end{equation}
 Note that $\|\bar{\mathbf{e}}(t)\|_1=\sum_{i=1}^M |\bar{e}_i(t)|$, combining \eqref{eq:boundbare} and \eqref{eq:boundrho0} yields for each $1\leq i\leq M$ and all $t\geq t_*$,
\begin{align}\label{eq:sumnormbound}
 \|\bar{\mathbf{e}}(t)\|_1 & \leq M( |z_{\infty}|+\epsilon)=\mu( |z_{\infty}|-\epsilon)\leq \mu |\bar{e}_i(t)|.
\end{align}

 On the other hand, by Proposition~\ref{pro:propertyL}, we have $\bar{l}_{ii}(t)=-\sum_{j\neq i} \bar{l}_{ij}(t)$, $\forall i$. Thus, one can obtain from \eqref{eq:errordynCSI} that  for each $1\leq i\leq M$,
\begin{align*}
\bar{e}_i(t+1)&=(1-a(t)\chi_i(t))\bar{e}_i(t)-a(t)\!\sum_{j\neq i} \bar{l}_{ij}(t)(\bar{e}_j(t)-\bar{e}_i(t))\\
&\leq (1-a(t)\chi_i(t))|\bar{e}_i(t)|+a(t)\zeta_i(t), \ \forall t\geq t_*,
\end{align*}
where $\zeta_i(t)=\sum_{j\neq i} |\bar{l}_{ij}(t)||\bar{e}_j(t)-\bar{e}_i(t)|$, and in the last inequality we use the fact that $1-a(t)\chi_i(t)\geq 0$, $\forall t\geq t_*$,  by \eqref{eq:boundrho0}. As a result, we can obtain for all $t\geq t_*$,
\begin{equation}\label{eq:limit0}
  \|\bar{\mathbf{e}}(t+1)\|_1\leq  \|\bar{\mathbf{e}}(t)\|_1-a(t)\sum_{i=1}^M \bigl[\chi_i(t)|\bar{e}_i(t)|-\zeta_i(t)\bigr],
\end{equation}

 Let $s_1,s_2,\dots$ be the time instants at which at least one SN has a measurement, then  $\sum_{i=1}^M \chi_i(s_k)\geq 1$, $\forall k\geq 1$, and $\chi_i(t)=0$, $\forall i$, $\forall s_k<t<s_{k+1}$.  Define $\zeta_{s_k,s_{k+1}}\triangleq \sum_{i=1}^M\sum_{s=s_k}^{s_{k+1}-1}a(s)\zeta_i(s)$, then summing up \eqref{eq:limit0} between $s_k$ and $s_{k+1}$ yields
\begin{align}
  \|\bar{\mathbf{e}}(s_{k+1})\|_1&\leq   \|\bar{\mathbf{e}}(s_{k})\|_1-a(s_k)\sum_{i=1}^M \chi_i(s_k) |\bar{e}_{i}(s_k)|+\zeta_{s_k,s_{k+1}}\notag\\
  &\leq\|\bar{\mathbf{e}}(s_{k})\|_1-a(s_k) \min_i|\bar{e}_{i}(s_k)|+\zeta_{s_k,s_{k+1}}\notag\\
  &\leq \left(1-\frac{a(s_k)}{\mu}\right)\|\bar{\mathbf{e}}(s_{k})\|_1+\zeta_{s_k,s_{k+1}},\label{eq:esk+1_iteration}
\end{align}
where  in the last step use was made of \eqref{eq:sumnormbound}.

 Consider the iteration \eqref{eq:esk+1_iteration}, we first note from \eqref{eq:boundrho0} that $0<\mu^{-1}a(s_k)<1$. And, by Assumption 2, one has $1\leq s_{k+1}-s_k\leq T$, which together with \eqref{eq:persistence2} implies  $\sum_{k=1}^{\infty} a(s_k)=\infty$.  Further, we know that $\mathbf{\bar{L}}(t)$ is bounded, since $\alpha_i(t)$ is bounded, and $\lim_{t\to \infty}\bar{\mathbf{e}}(t)=z_{\infty}\mathbf{1}$ by the Claim in step~1. It follows that $\lim_{t\to \infty}\zeta_i(t)=0$, $\forall 1\leq i\leq M$, which implies 
\begin{align*}
\lim_{k\to \infty}\frac{\zeta_{s_k,s_{k+1}}}{a(s_k)}&=\lim_{k\to \infty}\sum_{i=1}^M\sum_{s=s_k}^{s_{k+1}-1}\frac{a(s)}{a(s_k)}\zeta_i(s)\\
 &\leq T \lim_{k\to \infty}\sum_{i=1}^M \max_{t_k\leq s<t_{k+1}} \zeta_i(s)=0,
\end{align*}
 where in the last step we use the facts that $ s_{k+1}-s_k\leq T$, $\forall k$, and $a(t)$ is a nonincreasing function of $t$ by \eqref{eq:persistence}. 
 Hence, by Lemma~3 of \cite[p.45]{Pol87}, we infer from \eqref{eq:esk+1_iteration} that $\lim_{k\to \infty}\|\bar{\mathbf{e}}(s_{k})\|_1=0$.
 This reveals that the subsequence $\{\bar{e}_i(s_{k})\}_{k\geq 1}$ converges to 0 for each $i$. However, by the Claim in step 1,  $\lim_{t\to \infty}\bar{e}_i(t)=z_{\infty}\neq 0$, $\forall 1\leq i\leq M$, which is a contradiction. Consequently, it is necessary that  $z_{\infty}=0$. 

 Therefore, we have $\lim_{t\to \infty}\bar{\mathbf{e}}(t)=\mathbf{0}$, from which the theorem follows.

\section{Proof of Claim}\label{app:proofclaim}
 
  By Lemma~\ref{lem:propertiesofz}, we know that either $z_M(t)$ or $z_1(t)$ converges to a finite limit. We first consider the former case.  

  The following procedure will be directly operated on \eqref{eq:errordynnewscale} in the new time scale. 
   Recalling 
   the assumption that $\mathbb{E}\{\mathbf{\Gamma}(t)\}=\mathbf{0}$, we can derive from \eqref{eq:1stmoment} and \eqref{eq:transitionmatrix} that 
 \[
 \bar{\mathbf{e}}(t_{k+1})=\Biggl(\mathbf{I}-\!\sum_{s=t_k}^{t_{k+1}-1}\!a(s)(\mathbf{\Delta}(s)+\bar{\mathbf{L}}(s))+a^2(t_k)\bar{\mathbf{D}}_k\Biggr)\bar{\mathbf{e}}(t_k),
\]
 where $\bar{\mathbf{D}}_k$ is the matrix associated with the higher order terms of $a(t_k)$. 
It thus follows from Proposition~\ref{pro:propertyL} that  
\begin{equation}
\bar{e}_i(t_{k+1}) =\Biggl(1-\sum_{s=t_k}^{t_{k+1}-1} a(s)\chi_i(s)\Biggr)\bar{e}_i(t_k)-f_{i}(t_k,t_{k+1})+a^2(t_k) (\bar{\mathbf{D}}_k \bar{\mathbf{e}}(t_k))_i, \ 1\leq i\leq M, \label{eq:barei_exp}
\end{equation}
where 
\[
f_{i}(t_k,t_{k+1})=\sum_{s=t_k}^{t_{k+1}-1}\!\! a(s) \sum_{j\neq i}\bar{l}_{ij}(s) (\bar{e}_j(t_k)-\bar{e}_i(t_k)),
\]
 and $(\bar{\mathbf{D}}_k \bar{\mathbf{e}}(t_k))_i$ is the $i$-th component of $\bar{\mathbf{D}}_k \bar{\mathbf{e}}(t_k)$.

  Under the assumptions, we know that $\alpha_i(t)$ is bounded and $\bar{h}_{ij}=\mathbb{E}\{h_{ij}(t)\}$, $\forall i,j$. Moreover, $a(t)$ is nonincreasing in $t$ by \eqref{eq:persistence}, and $t_{k+1}-t_k=\tau$. Hence, we can find a constant $c_1>0$ such that
  \begin{equation}\label{eq:boundLD}
  \max\{\|\mathbf{\bar{L}}(t)\|_\infty, \|\bar{\mathbf{D}}_k\|_\infty\}\leq c_1, \ \forall t, k,
  \end{equation}
  which together with Lemma~\ref{lem:propertiesofz} implies
\begin{equation}\label{eq:Dei_Lbound}
(\bar{\mathbf{D}}_k \bar{\mathbf{e}}(t_k))_i\geq -\|\bar{\mathbf{D}}_k \bar{\mathbf{e}}(t_k)\|_\infty\geq -c_0 c_1, \ \forall i.
\end{equation} 

  We use the induction method to proceed with the proof.  

\textbf{Step 1:} (Initial step) We have $\lim_{t\to\infty}z_M(t)=z_{M\infty}\leq 0$ by Lemma~\ref{lem:propertiesofz}.

\textbf{Step 2:} (Inductive step)
  Assume that for some $1<l\leq M$, 
\begin{equation}\label{eq:induction0}
\lim_{t\to\infty} z_i(t)=z_{M\infty}, \ \forall l\leq i\leq  M.
\end{equation}
 We next show that \eqref{eq:induction0} holds for $i=l-1$.

 \textbf{Step 2.1:} (Contradiction argument) Suppose it is not the case, i.e., $z_{l-1}(t) \nrightarrow z_{M\infty}$,  as $t\to \infty$. By the ordering of $\{z_i(t)\}_{1\leq i\leq M}$, we have $\liminf_{t\to\infty} z_{l-1}(t)\geq \lim_{t\to\infty}z_M(t) =z_{M\infty}$. As a result, there exists a constant $\epsilon_0>0$ and a subsequence of $\{t_k\}_{k\geq 0}$, denoted with abuse of notation  by $\{t_k\}_{k\geq 0}$ again, such that 
 \begin{equation}\label{eq:contradiction0}
 z_{l-1}(t_k)\geq z_{M\infty}+\epsilon_0,  \ \text{for large } \ k.
 \end{equation}

 By \eqref{eq:persistence} and \eqref{eq:induction0}, for any $0<\epsilon_1<\epsilon_0/3$, we can find a $t_k^*\geq 0$ such that
 \begin{equation}\label{eq:induction}
  a(t)\leq \epsilon_1,  \ \ |z_i(t)-z_{M\infty}|\leq \epsilon_1, \ \forall l\leq i\leq  M, \forall t\geq t_{k^*},
\end{equation}
which implies 
\begin{equation}\label{eq:asquareUbound}
 a^2(t_k) \leq\epsilon_1\sum_{s=t_k}^{t_{k+1}-1}a(s), \ \forall t_k\geq t_{k^*}.
\end{equation} 
 Further, for this $t_k^*$, it follows form \eqref{eq:contradiction0} that 
  \begin{equation}\label{eq:contradictionLimit}
   z_{i}(t_{k^*})\geq z_{M\infty}+\epsilon_0, \ \forall 1\leq i\leq  l-1,
 \end{equation}
since $z_1(t_{k^*})\geq z_2(t_{k^*})\geq \dots\geq z_{l-1}(t_{k^*})$.

 Associate the states $z_l(t_{k^*}), \dots, z_M(t_{k^*})$ at time $t_{k^*}$ with the SNs $l_{t_{k^*}},\dots, M_{t_{k^*}}$, respectively.  Denote $\mathcal{I}_{k^*}\triangleq \{l_{t_{k^*}},\dots,M_{t_{k^*}}\}$ and $\mathcal{I}_{k^*}^c\triangleq \mathcal{I}_s\backslash \mathcal{I}_{k^*}$. 

 \textbf{Step 2.2:} (Evolution of the states in $\mathcal{I}_{k^*}^c$) Consider arbitrary element $i\in\mathcal{I}_{k^*}^c$,  we derive from Lemma~\ref{lem:propertiesofz} and \eqref{eq:boundLD} that
$|f_{i}(t_k,t_{k+1})|\leq 2c_0c_1\sum_{s=t_k}^{t_{k+1}-1} a(s)$ and $|\bar{e}_i(t)|\leq c_0$. Let $c_{\epsilon_0}\triangleq c_0(1+2c_1+c_1\epsilon_0/3)$, it then follows from \eqref{eq:barei_exp} that
\begin{align}
  \hspace{-0.2em} \bar{e}_i(t_{k+1})&\overset{(a)}{\geq}\min_{i\in \mathcal{I}_{k^*}^c} \bar{e}_i(t_k)-c_0(1+2c_1)\!\sum_{s=t_k}^{t_{k+1}-1}\!\! a(s)-c_0c_1 a^2(t_k)\notag\\
   &\overset{(b)}{\geq} \min_{i\in \mathcal{I}_{k^*}^c} \bar{e}_i(t_k)-c_0(1+2c_1+c_1\epsilon_1)\sum_{s=t_k}^{t_{k+1}-1} a(s)\notag\\
   &\overset{(c)}{\geq} \min_{i\in \mathcal{I}_{k^*}^c} \bar{e}_i(t_k)-c_{\epsilon_0}\sum_{s=t_k}^{t_{k+1}-1} a(s), \ \forall t_{k} \geq t_{k^*}, \label{eq:summin}
\end{align}
 where  $(a)$ is due to \eqref{eq:Dei_Lbound}, $(b)$ follows from \eqref{eq:asquareUbound}, and $(c)$ is due to the fact that $\epsilon_1<\epsilon_0/3$.

By \eqref{eq:persistence2}, we can find an integer $K^*$ depending on $\epsilon_0$ and $t_{k^*}$ so that
\begin{equation}\label{eq:boundsumrho}
\frac{\kappa\epsilon_0}{3c_{\epsilon_0}}\leq \sum_{s=t_{k^*}}^{t_{K^*}-1}a(s)\leq \frac{\epsilon_0}{3c_{\epsilon_0}}.
\end{equation}
Hence, summing up the inequalities \eqref{eq:summin} from $t_{k^*}$ to arbitrary $t_k\leq {t_{K^*}}$  yields
 \begin{align*}
 \min_{i\in \mathcal{I}_{k^*}^c}\bar{e}_i(t_{k})&\geq \min_{i\in \mathcal{I}_{k^*}^c} \bar{e}_i(t_{k^*})-c_{\epsilon_0}\sum_{s=t_{k^*}}^{t_{K^*}-1}a(s)\notag\\
 &\overset{(a)}{\geq} z_{l-1}(t_{k^*})-\frac{\epsilon_0}{3}\notag\\
 &\overset{(b)}{\geq} z_i(t_k)+\frac{\epsilon_0}{3}, \ \forall l\leq i\leq M,
\end{align*}
where $(a)$ follows from the ordering of $\{z_i(t)\}_{1\leq i\leq M}$ and \eqref{eq:boundsumrho}, and $(b)$ is due to \eqref{eq:induction} and \eqref{eq:contradictionLimit}. This means that
\begin{equation}\label{eq:greatestValue}
 \min_{i\in \mathcal{I}_{k^*}^c} \bar{e}_i(t_k)\geq \max_{j\in \mathcal{I}_{k^*}}\bar{e}_j(t_k)+\frac{\epsilon_0}{3},\ \forall t_{k^*}\leq t_k\leq t_{K^*}.
\end{equation}

\textbf{Step 2.3:} (Evolution of the states in $\mathcal{I}_{k^*}$)  Consider arbitrary $j\in \mathcal{I}_{k^*}$ and  $t_{k^*}\leq t_k\leq t_{K^*}$.  
Since $\mathcal{I}_{k^*}\cup \mathcal{I}_{k^*}^c=\mathcal{I}_s$, one has
\begin{align*}
 f_j(t_k,t_{k+1})&=\sum_{s=t_k}^{t_{k+1}-1}a(s)\sum_{p\in \mathcal{I}_{k^*}\backslash\{j\}}\bar{l}_{jp}(s)(\bar{e}_p(t_k)-\bar{e}_j(t_k))+\sum_{s=t_k}^{t_{k+1}-1}a(s)\sum_{p\in \mathcal{I}_{k^*}^c}\bar{l}_{jp}(s)(\bar{e}_p(t_k)-\bar{e}_j(t_k))\\
 &\triangleq f_j^1+f_j^2.
\end{align*}
 Considering $f_j^1$, we derive from \eqref{eq:boundLD} and \eqref{eq:induction} that
\begin{equation}\label{eq:fj1}
f_j^1\leq 2c_1\epsilon_1\sum_{s=t_k}^{t_{k+1}-1}a(s).
\end{equation}
As for $f_j^2$, by Proposition~\ref{pro:propertyL}, we know that $\bar{l}_{ij}(t)\leq 0$, $\forall j\neq i$. It then follows from \eqref{eq:greatestValue} that
\begin{equation}\label{eq:fj2}
f_j^2\leq \frac{\epsilon_0}{3}\sum_{s=t_k}^{t_{k+1}-1} a(s)\sum_{p\in\mathcal{I}_{k^*}^c}\bar{l}_{jp}(s).
\end{equation}
Define $g_j(t_{k})\triangleq \bar{e}_j(t_{k})-(1- \sum_{s=t_{k^*}}^{t_{k}-1}a(s)\chi_j(s))z_{M\infty}$.
We can derive from \eqref{eq:barei_exp}, \eqref{eq:fj1} and \eqref{eq:fj2} that
\begin{align*}
g_j(t_{k+1})&=g_j(t_k)+\sum_{s=t_k}^{t_{k+1}-1}a(s)\chi_j(s)(z_{M\infty}-\bar{e}_j(t_k))-f_j(t_k,t_{k+1})+a^2(t_k)(\bar{\mathbf{D}}_k\bar{\mathbf{e}}(t_k))_j\notag\\
&\geq g_j(t_k)-\sum_{s=t_k}^{t_{k+1}-1}a(s)\Bigl(c_2\epsilon_1+\frac{\epsilon_0}{3}\sum_{p\in\mathcal{I}_{k^*}^c}\bar{l}_{jp}(s)\Bigr),
\end{align*}
where in the last line use was made of \eqref{eq:Dei_Lbound}, \eqref{eq:induction}, \eqref{eq:asquareUbound}, and we denote $c_2\triangleq 1+2c_1+c_0c_1$.  Let $g_{\text{sum}}({t_k})\triangleq \sum_{j\in\mathcal{I}_{k^*}}g_j(t_{k})$, then it follows that 
\begin{equation}\label{eq:evolleast}
g_{\text{sum}}({t_{k+1}})\geq g_{\text{sum}}({t_k})-Mc_2\epsilon_1\sum_{s=t_k}^{t_{k+1}-1}a(s)
-\frac{\epsilon_0}{3}\sum_{s=t_k}^{t_{k+1}-1}a(s)\sum_{j\in\mathcal{I}_{k^*},p\in\mathcal{I}_{k^*}^c}\bar{l}_{jp}(s).
\end{equation}

Under Assumption 1, we can find some $t_k\leq t_{k_{0}}< t_{k+1}$ such that there exists either one edge $(p_0,j_0)\in\mathcal{E}(t_{k_{0}})$ or one path $p_0\to q_0\to j_0$, where $p_0\in\mathcal{I}_{k^*}^c$, $j_0\in\mathcal{I}_{k^*}$ and $q_0\in\mathcal{I}_r$. 
Since $\alpha_i(t)$, $\forall i$, is bounded,  it follows from Proposition~\ref{pro:propertyL} that
\begin{equation}
\sum_{s=t_k}^{t_{k+1}-1}a(s)\sum_{j\in\mathcal{I}_{k^*},p\in\mathcal{I}_{k^*}^c}\bar{l}_{jp}(s)\leq a(t_{k_0})\bar{l}_{j_0p_0}(s)
\leq -\frac{c_3}{c_4} \sum_{s=t_k}^{t_{k+1}-1}a(s),\label{eq:sum_at_Lbound}
\end{equation}
where $c_3=\inf_t \alpha_j(t)\min\{\bar{h}_{ij}, \inf_t\alpha_q(t)\bar{h}_{ip}\bar{h}_{pj}\}$, $c_4=(\underline{c}^{-\tau}-1)/(1-\underline{c})+\tau$, and in the last line we use \eqref{eq:persistence} and $t_{k+1}-t_k=\tau$ to obtain $\sum_{s=t_k}^{t_{k+1}-1} a(s)/a(t_{k_0})\leq (\underline{c}^{t_{k}-t_{k_0}}-1)(1-\underline{c})+\tau\leq  c_4$.
By iterating \eqref{eq:evolleast} and using \eqref{eq:sum_at_Lbound}, we arrive at the following relation
\begin{equation}\label{eq:contradiction1}
 g_{\text{sum}}({t_{K^*}})\geq g_{\text{sum}}({t_{k^*}})+\frac{c_3\epsilon_0-3Mc_2c_4\epsilon_1}{3c_4}\sum_{s=t_{k^*}}^{t_{K^*}-1}a(s).
\end{equation}

\textbf{Step 2.4:} (A contradiction)
Choose $\epsilon_1<\min\bigl\{\epsilon_0/3,c_3 \kappa \epsilon_0^2/(3M c_4(\kappa c_2 \epsilon_0+6 c_{\epsilon_0}))\bigr\} $,  then one has $c_3\epsilon_0-3Mc_2c_4\epsilon_1>0$, and  the lower bound in \eqref{eq:boundsumrho} enables \eqref{eq:contradiction1} to be
\begin{equation}\label{eq:contradiction2}
 g_{\text{sum}}({t_{K^*}})> g_{\text{sum}}({t_{k^*}})+2M\epsilon_1.
\end{equation}

 On the other hand, it follows from \eqref{eq:induction} that $g_j(t_{k^*})=\bar{e}_j(t_{k^*})-z_{M\infty}\geq -\epsilon_1$.
Moreover, recalling that $z_{M\infty}\leq 0$ by Lemma~\ref{lem:propertiesofz}, we have 
$g_j(t_{K^*})\leq \bar{e}_j(t_{K^*})-z_{M\infty}\leq \epsilon_1$.
 Combining the previous two inequalities yields
\begin{equation}
 g_{\text{sum}}(t_{K^*})-g_{\text{sum}}(t_{k^*})\leq 2\sum_{j\in\mathcal{I}_{k^*}} \epsilon_1\leq 2M\epsilon_1,
\end{equation}
which is a contradiction to \eqref{eq:contradiction2}. Therefore, the relation \eqref{eq:induction0} holds for $i=l-1$.

\textbf{Step 3:} (Conclusion)
 By induction, we conclude that \eqref{eq:induction0} holds for all $1\leq i\leq M$. Remember that $z_M(t)\leq \bar{e}_i(t)\leq z_1(t)$, $\forall i$. Hence, the original state sequences $\{\bar{e}_i(t)\}_{t\geq 0}$ satisfy $\lim_{t\to \infty}\bar{e}_i(t)=z_{M\infty}$, $\forall i$. 

 The proof for the case that  $\lim_{t\to\infty}z_1(t)=z_{1\infty}$ is quite similar, so we omit the details. 
 This completes the proof.

\section{Proof of Theorem~\ref{thm:consistency}}\label{app:proofconsistency}
For  notational simplicity, we use $\mathbf{\Psi}_{k,s}$ and $\mathbf{\Psi}_{k}$ for the transition matrices $\mathbf{\Psi}_{t_{k+1},s}$ and $\mathbf{\Psi}_{t_{k+1},t_k}$, respectively. 
   Using \eqref{eq:transitionmatrix}, we can express $\mathbf{\Psi}_{k,s}$ and $\mathbf{\Psi}_{k}$  as 
   \begin{equation}\label{eq:transitionmatrixshort}
   \mathbf{\Psi}_{k,s}=\mathbf{I}-\mathbf{Q}_{k,s}+\mathbf{R}_{k,s},\  \mathbf{\Psi}_{k}=\mathbf{I}-\mathbf{Q}_{k}+\mathbf{R}_{k},
   \end{equation}
   where $\mathbf{Q}_{k,s}=\sum_{j=s}^{t_{k+1}-1}a(j)\mathbf{\Phi}(j)$,  $\mathbf{R}_{k,s}$ collects all the high order terms with respect to $a(t_k)$, and $\mathbf{Q}_{k}$, $\mathbf{R}_{k}$ are short for $\mathbf{Q}_{k,t_{k}}$, $\mathbf{R}_{k,t_k}$, respectively.  The proof consists of two steps:

\textbf{Step 1:} (Convergence of $V(t_{k})$) We derive from \eqref{eq:errordynnewscale} that
\begin{align}
\mathbb{E}\left\{V(t_{k+1})|\mathcal{F}_{t_k}\right\}
&=\mathbf{e}^T (t_k) \mathbb{E}\{\mathbf{\Psi}_{k}^T\mathbf{\Psi}_{k}\}\mathbf{e}(t_k)+2\mathbf{e}^T(t_k)\mathbb{E}\{\mathbf{\Psi}_{k}^T\boldsymbol{\phi}(t_k)\}+\mathbb{E}\{\|\boldsymbol{\phi}(t_k)\|_2^2\}\notag\\
&\triangleq V_1(k)+V_2(k)+V_3(k),\label{eq:Vk+1}
\end{align}
where $\boldsymbol{\phi}(t_k)=\sum_{s=t_k}^{t_{k+1}-1}a(s)\mathbf{\Psi}_{k,s+1} (\theta \mathbf{\Gamma}(s)\mathbf{1}+\mathbf{\Delta}(s) \mathbf{w}(s)+\mathbf{v}(s))$.

 In the following, we will establish upper bounds of  $V_i(k)$, $i=1,2,3$, respectively.

 \emph{(i) Bound of $V_1(k)$.}   In view of $\mathbb{E}\{\mathbf{\Gamma}(t)\}=\mathbf{0}$,  we have $\mathbb{E}\{\mathbf{Q}_k+\mathbf{Q}^T_k\}=\mathbf{J}_k$. It thus follows from \eqref{eq:transitionmatrixshort} that
\begin{align}
 V_1(k)&\leq \mathbf{e}^T(t_k) (\mathbf{I}-\mathbf{J}_k)\mathbf{e}(t_k)+\bigl(2\|\mathbb{E}\{\mathbf{R}_{k}\}\|+\|\mathbb{E}\{\mathbf{Q}_{k}^T\mathbf{Q}_{k}\}\| +2\|\mathbb{E}\{\mathbf{Q}_{k}^T\mathbf{R}_{k}\}\|
 +\|\mathbb{E}\{\mathbf{R}_{k}^T\mathbf{R}_{k}\}\| \bigr)V(t_k)\notag\\
 &\leq \mathbf{e}^T(t_k) (\mathbf{I}-\mathbf{J}_k)\mathbf{e}(t_k)+\Bigl(2\sqrt{\mathbb{E}\{\|\mathbf{R}_{k}\|^2\}}+\bigl(\sqrt{\mathbb{E}\{\|\mathbf{Q}_{k}\|^2\}}+\sqrt{\mathbb{E}\{\|\mathbf{R}_{k}\|^2\}}\bigr)^2\Bigr)V(t_k),\label{eq:V10}
\end{align}
 where in the last line we use the Cauchy-Schwarz inequality $\mathbb{E}\{|x||y|\}\leq \sqrt{\mathbb{E}\{x^2\}}\sqrt{\mathbb{E}\{y^2\}} $, $\forall x,y\in \mathbb{R}$ \cite[p.130]{Gut13}. 
 
 Note that $a(t)$ is nonincreasing by \eqref{eq:persistence} and $t_{k+1}-t_k=\tau$, then applying the $c_r$ inequality \cite[p.127]{Gut13} to $\|\mathbf{R}_{k}\|^2$  yields
 \begin{equation}\label{eq:Rk2bound}
 \mathbb{E}\{\|\mathbf{R}_{k}\|^2\}\leq 2^{\tau} a^4(t_k) \sum_{s_1>s_2} \mathbb{E}\{\|\mathbf{\Phi}(s_1)\|^2\} \mathbb{E}\{\|\mathbf{\Phi}(s_2)\|^2\}
 +\dots+ 2^{\tau} a^{2\tau}(t_k)\prod_{s=t_k}^{t_{k+1}-1} \mathbb{E}\{\|\mathbf{\Phi}(s)\|^2\}.
 \end{equation}
 Similarly, one obtains
 \begin{equation}\label{eq:Qk2bound}
\mathbb{E}\{\|\mathbf{Q}_{k}\|^2\}\leq \tau a^2(t_k)\sum_{s=t_k}^{t_{k+1}-1} \mathbb{E}\{\|\mathbf{\Phi}(s)\|^2\}.
 \end{equation}
 Since $\mathbb{E}\{\|\mathbf{\Phi}(s)\|^2\}\leq \text{trace}(\mathbb{E}\{\mathbf{\Phi}(s)^T\mathbf{\Phi}(s)\})$, it follows from the definition of  $\mathbf{\Phi}(s)$ (see Table~\ref{tab:notation}) that 
 \[
 \mathbb{E}\{\|\mathbf{\Phi}(s)\|^2\}\leq \sum_{j=1}^M (\chi_i(s)+b_i(s))^2+\!\sum_{1\leq j\neq i\leq M}\! \mathbb{E}\{l_{ij}(s)^2\}<\infty,
 \]
 which is bounded under Assumption 3 and the assumptions that $\alpha_i(t)$, $b_i(t)$, $\forall i$, are bounded. Substituting this bound into \eqref{eq:Rk2bound}, \eqref{eq:Qk2bound} and then into \eqref{eq:V10} implies
\begin{equation}\label{eq:V1bound}
V_1(k)\leq \mathbf{e}^T(t_k) (\mathbf{I}-\mathbf{J}_k)\mathbf{e}(t_k) +\mathcal{O}(a^2(t_k)) V(t_k).
\end{equation}

\emph{(ii) Bound of $V_2(k)$.} Under the independence assumptions, we have
 \begin{align}
  V_2(k)&=2\theta \mathbf{e}^T(t_k) \sum_{s=t_k}^{t_{k+1}-1} a(s)\mathbb{E}\{\mathbf{\Psi}_{k}^T \mathbf{\Psi}_{k,s+1} \mathbf{\Gamma}(s)\}\mathbf{1}\notag\\
  &\overset{(a)}{\leq}  \theta \sum_{s=t_k}^{t_{k+1}-1}\bigl( a^2(s) V(t_k)+\|\mathbb{E}\{\mathbf{\Psi}_{k}^T \mathbf{\Psi}_{k,s+1} \mathbf{\Gamma}(s)\}\mathbf{1}\|_2^2\bigr)\notag\\
  &\overset{(b)}{\leq} \theta  \tau a^2(t_k) V(t_k)+ \theta M\sum_{s=t_k}^{t_{k+1}-1}\!\|\mathbb{E}\{\mathbf{\Psi}_{k}^T \mathbf{\Psi}_{k,s+1} \mathbf{\Gamma}(s)\}\|^2,\label{eq:V2bound0}
 \end{align}
  where $(a)$ follows from the inequality   $2\mathbf{z}_1^T\mathbf{z}_2\leq \|\mathbf{z}_1\|_2^2+\|\mathbf{z}_2\|_2^2$, for any two vectors $\mathbf{z}_1,\mathbf{z}_2$, and $(b)$ is due to \eqref{eq:persistence} and the fact that $t_{k+1}-t_k=\tau$. 

  Recall that $\mathbb{E}\{\mathbf{\Gamma}(t)\}=\mathbf{0}$, $\forall t$, and the independence assumption, we have $\mathbb{E}\{(\mathbf{Q}_{k,s+1}-\mathbf{R}_{k,s+1}) \mathbf{\Gamma}(s)\}=0$. Hence,  we can derive from \eqref{eq:transitionmatrixshort} that
  \begin{equation*}
\|\mathbb{E}\{\mathbf{\Psi}_{k}^T \mathbf{\Psi}_{k,s+1} \mathbf{\Gamma}(s)\}\|\leq \bigl(1+\mathbb{E}\{\|\mathbf{Q}_{k,s+1}-\mathbf{R}_{k,s+1}\|\}\bigr)\mathbb{E}\{\|\mathbf{Q}_{k}-\mathbf{R}_{k}\| \|\mathbf{\Gamma}(s)\|\}.
  \end{equation*}
  Referring to Table~\ref{tab:notation}, we have 
\begin{equation}\label{eq:Gammabound}
  \mathbb{E}\{\|\mathbf{\Gamma}(s)\|^2\}=\max_i \mathbb{E}\bigl\{(l_{ii}(s)-b_i(s))^2\bigr\}<\infty,
\end{equation}
  which is bounded under Assumption~3. Further, following the same arguments as in \eqref{eq:Rk2bound} and \eqref{eq:Qk2bound}, we can show that $\mathbb{E}\{\|\mathbf{Q}_{k,s}-\mathbf{R}_{k,s}\|^2\}\leq 2\mathbb{E}\{\|\mathbf{Q}_{k,s}\|^2+\|\mathbf{R}_{k,s}\|^2\}=\mathcal{O}(a^2(t_k))$. Therefore, employing the Cauchy-Schwarz inequality \cite[p.130]{Gut13} implies that $\|\mathbb{E}\{\mathbf{\Psi}_{k}^T \mathbf{\Psi}_{k,s+1} \mathbf{\Gamma}(s)\}\|^2$ is of the order $\mathcal{O}(a^2(t_k))$. This together with \eqref{eq:V2bound0} shows that 
\begin{equation}\label{eq:V2bound}
V_2(k)\leq \theta  \tau a^2(t_k) V(t_k)+ \mathcal{O}(a^2(t_k)).
\end{equation}

 \emph{(iii) Bound of $V_3(k)$.}  One can get
 \begin{align*}
 V_3(k)&\overset{(a)}{=} \sum_{s=t_k}^{t_{k+1}-1} a^2(s)\Bigl(\theta^2 \mathbb{E}\{\|\mathbf{\Psi}_{k,s+1}\mathbf{\Gamma}(s)\mathbf{1}\|_2^2\}+\mathbb{E}\{\|\mathbf{\Psi}_{k,s+1}\mathbf{\Delta}(s)\mathbf{w}(s)\|_2^2\}+\mathbb{E}\{\|\mathbf{\Psi}_{k,s+1}\mathbf{v}(s)\|_2^2\}\Bigr.\notag\\
 &\Bigl.\relphantom{\leq}{} +2\theta \mathbb{E}\{\mathbf{1}^T \mathbf{\Gamma}(s)\mathbf{\Psi}_{k,s+1}^T\mathbf{\Psi}_{k,s+1} \mathbf{v}(s)\}\Bigr)\notag\\
 &\overset{(b)}{\leq} a^2(t_k) \sum_{s=t_k}^{t_{k+1}-1} \mathbb{E}\{\|\mathbf{\Psi}_{k,s}\|^2\} \Bigl(\mathbb{E}\{\|\mathbf{w}(s)\|^2\}+\bigl(\theta  \sqrt{M\mathbb{E}\{\|\mathbf{\Gamma}(s)\|^2\}}+ \sqrt{\mathbb{E}\{\|\mathbf{v}(s)\|^2\}}\bigr)^2 \Bigr),
 \end{align*}
 where $(a)$ follows from the Assumptions~2-4, and $(b)$ is due to \eqref{eq:persistence} and the  Cauchy-Schwarz inequality \cite[p.130]{Gut13}. 

 Under Assumptions 2-4, we know that $\mathbb{E}\{\|\mathbf{w}(s)\|^2\}$ and $\mathbb{E}\{\|\mathbf{v}(s)\|^2\}$ are bounded. Repeating the same arguments as in \eqref{eq:Rk2bound} and \eqref{eq:Qk2bound}, we can show that $\mathbb{E}\{\|\mathbf{\Psi}_{k,s}\|^2\}$ is bounded. This together \eqref{eq:Gammabound} implies that
\begin{equation}\label{eq:V3bound}
 V_3(k)=\mathcal{O}(a^2(t_k)).  
\end{equation} 

Substituting \eqref{eq:V1bound}, \eqref{eq:V2bound} and \eqref{eq:V3bound} into \eqref{eq:Vk+1} yields
\begin{equation}\label{eq:RobSeg}
 \mathbb{E}\{V(t_{k+1})|\mathcal{F}_{t_k}\}\leq \Bigl(1+(\tau \theta+\mathcal{O}(1)) a^2(t_k)\Bigr)V(t_k)\\
 -\mathbf{e}^T(t_k) \mathbf{J}_k \mathbf{e}(t_k)+\mathcal{O}(a^2(t_k)),
\end{equation}
 where $\mathcal{O}(1)$ denotes a positive constant. Applying the Robbin-Siegmund theorem \cite[p.50]{Pol87} to \eqref{eq:RobSeg} ensures that there is a random variable $\eta$ such that 
\begin{equation}\label{eq:RobSegresult}
 V(t_k)\to \eta, \ \text{w. p. 1}, \ \text{and}\  \  \sum_{k=0}^{\infty} \mathbf{e}^T(t_k) \mathbf{J}_k \mathbf{e}(t_k)< \infty.
\end{equation}
 However, by \eqref{eq:boundcondconsis}, we have 
\begin{equation}\label{eq:RobSeglowerbound}
\mathbf{e}^T(t_k)\mathbf{J}_k \mathbf{e}(t_k)\geq c^*a(t_k) V(t_k),\  \forall k\geq 0.
\end{equation}
Recalling that $t_{k+1}-t_k=\tau$, it follows from \eqref{eq:persistence2} that $\sum_{k=0}^{\infty} a(t_k)=\infty$. Combining \eqref{eq:RobSegresult} and \eqref{eq:RobSeglowerbound} implies that $\eta=0$. Therefore, 
we have  $\lim_{k\to \infty}\mathbb{E}\{V(t_k)\}=0$.

\textbf{Step 2:} (Passing from the convergence of $\mathbb{E}\{V(t_k)\}$ to the convergence of $\mathbb{E}\{V(t)\}$)
For any $t\geq \tau$, define $\tau_t\triangleq \tau\lfloor t/\tau-1\rfloor$, where $\lfloor t/\tau-1\rfloor$ denotes the largest integer no greater than $t/\tau-1$, then $\tau\leq t-\tau_t\leq 2\tau$. This means that $\mathcal{G}_{[\tau_t,t)}$ is strongly connected on average by Assumption 1. Hence it can be shown that $\sum_{s=\tau_t}^{t}a(s)(2\mathbf{\Delta}(s)+\bar{\mathbf{L}}(s)+\bar{\mathbf{L}}^T(s))$ possesses a similar lower bound on its smallest eigenvalue as $\mathbf{J}_k$.
Similar to \eqref{eq:RobSeg}, and noting that $a(t)$ is nonincreasing by \eqref{eq:persistence}, we have
\[
 \mathbb{E}\{V(t+1)\}\leq \Bigl(1+\theta\sum_{s=\tau_t}^{t}a^2(s)\Bigr)\mathbb{E}\{V(\tau_t)\}+\mathcal{O}(a^2(\tau_t)),
\]
from which one knows that 
\[
\lim_{t\to \infty}\mathbb{E}\{V(t)\}= \lim_{t\to \infty}\mathbb{E}\{\|\mathbf{x}(t)-\theta \mathbf{1}\|^2\}=0,
\]
since $\lim_{t\to \infty}\mathbb{E}\{V(\tau_t)\}=0$ by step 1, and $\sum_{t=0}^{\infty} a^2(t)<\infty$ by \eqref{eq:conditionconsis}. This proves  the consistency of the estimates sequence $\{x_i(t)\}_{1\leq i\leq M, t\geq 0}$. The proof is thus complete.

\section{Proof of Proposition~\ref{pro:boundmaxeigenvalue}}\label{app:proofboundmaxeigenvalue}
 We start the proof by leveraging on the following
 lemma. 

\begin{lemma}[\cite{LuNg04}]\label{lem:spectralbound}
 For any irreducible nonnegative matrix $\mathbf{B}\in\mathbb{R}^{M\times M}$, let $P_S(\varsigma)\triangleq \mathbf{B}_{S^c}+\mathbf{B}_{S^c,S}(\varsigma I-\mathbf{B}_{S})^{-1}\mathbf{B}_{S,S^c}$, $\forall \varsigma>r_{\max}(\mathbf{B}_{S})$,  be the generalized Perron complement of $\mathbf{B}_{S}$. Then  we have
 $\rho(\mathbf{B})\leq \max\{\varsigma,r_{\max}(P_S(\varsigma))\}$
 and $r_{\max}(P_S(\varsigma))\leq \max_{j}\{r_j(\mathbf{B}_{S^c})+\upsilon(\varsigma)r_j(\mathbf{B}_{S^c,S})\}$, where $\upsilon(\varsigma)=\max_{i}(r_{i}(\mathbf{B}_{S,S^c})/(\varsigma-r_{i}(\mathbf{B}_{S})))$.
\end{lemma}

\emph{Proof of Lemma~\ref{pro:boundmaxeigenvalue}:}
  Since $\mathbf{B}_S$ is nonnegative, we have $\rho(\mathbf{B})\leq r_{\max}(\mathbf{B})$ \cite[p.24]{Minc88}. As a result,  $\varsigma \mathbf{I}-\mathbf{B}_S$ is a nonsingular M-matrix for any $\varsigma>r_{\max}(\mathbf{B}_S)$ \cite[p.159]{Minc88}.
  Hence, applying the Neumann expansion implies $P_S(\varsigma)=\sum_{j=0}^{\infty} {\varsigma}^{-j-1}\mathbf{B}_{S^c,S}\mathbf{B}_S^j \mathbf{B}_{S,S^c}$, $\forall \varsigma>r_{\max}(\mathbf{B}_S)$.
  Since $\mathbf{B}$ is irreducible, the associated graph $\mathcal{G}(\mathbf{B})$ is strongly connected \cite[p.78]{Minc88}. Thus there is a directed path $\mathfrak{n}_1\rightarrow\mathfrak{n}_{21}\rightarrow \dots\rightarrow \mathfrak{n}_{2j_0}\rightarrow\mathfrak{n}_3$ in $\mathcal{G}(\mathbf{B})$, where $\mathfrak{n}_1,\mathfrak{n}_3\in S^c$, $\mathfrak{n}_{21},\dots,\mathfrak{n}_{2j_0}\in S$, which implies that the $(\mathfrak{n}_1,\mathfrak{n}_3)$-th entry of the matrix $\mathbf{B}_{S^c,S}\mathbf{B}_S^{j_0-1}\mathbf{B}_{S,S^c}$ is positive. Consequently, $r_{\max}(P_S(\varsigma))$ is strictly decreasing in $\varsigma$ on the interval $(r_{\max}(\mathbf{B}_S),+\infty)$.

  By Lemma~\ref{lem:spectralbound}, we can see that the tightest upper bound of $\rho(\mathbf{B})$ is obtained at the point $\varsigma_*$ when $r_{\max}(\mathbf{B}_S)<\varsigma_*=r_{\max}(P_S(\varsigma_*))$. Since $r_{\max}(P_S(\varsigma))\leq \max_{j}\{r_j(\mathbf{B}_{S^c})+\upsilon(\varsigma)r_j(\mathbf{B}_{S^c,S})\}$, it follows that for some $i_0$ and $j_0$,
  \begin{equation}
  \varsigma_*^2- r_{i_0j_0}^+\varsigma_*+r_{j_0}(\mathbf{B}_{S^c})r_{i_0}(\mathbf{B}_S)-r_{i_0j_0}\leq 0.
  \end{equation}
  Hence, one obtains 
  \begin{equation}\label{eq:boundPerr}
 \varsigma_*\leq  \varsigma_{i_0j_0}\triangleq \frac{1}{2}\Bigl(r_{i_0j_0}^{+}+\sqrt{(r_{i_0j_0}^{-})^2+4r_{i_0j_0}}\Bigr).
  \end{equation}
  For each pair $i,j$, one has $0\leq r_{ij}\leq r_{\max}^2(\mathbf{B})-r_{\max}(\mathbf{B})r_{ij}^++r_{i}(\mathbf{B}_S)r_{j}(\mathbf{B}_{S^c})$, which shows that $(r_{ij}^-)^2 \leq (r_{ij}^-)^2+4r_{ij}\leq \bigl(r_{ij}^{+}-2r_{\max}(\mathbf{B})\bigr)^2$. This gives $r_i(\mathbf{B}_S)\leq\varsigma_{ij}\leq r_{\max}(\mathbf{B})$.
  Substituting the above bound into \eqref{eq:boundPerr} yields $\varsigma_*\leq \max_{i, j} \varsigma_{ij}\leq r_{\max}(\mathbf{B})$. This completes the proof.

\section{Proof of Proposition \ref{pro:weight} }\label{app:proofweight}
  First, it is noted that $a(t)=t^{-\alpha}$ is decreasing and tends to 0 for any $\alpha>0$. And $a(t+1)/a(t)=\left(1+1/t\right)^{-\alpha}\geq (2/3)^{\alpha}$,  $\forall t\geq 2$. Thus condition \eqref{eq:persistence} is satisfied.

  To prove that condition \eqref{eq:persistence2} is also satisfied. We first show that $T(t,\nu)$ is well-defined. Indeed, we have $t^{1-\alpha}-\alpha\kappa\nu\geq 2^{1-\alpha}-\alpha\kappa\nu>0$, $\forall t\geq 2$.
  Define the function $f(t)\triangleq (1-\alpha2^{\alpha-1}\kappa\nu)(\nu(t-1)^{\alpha}-1)-\kappa\nu t^{\alpha}$ on the interval $[0,\infty)$.  Then we have
\begin{align*}
  \lim_{t\to \infty} f(t)& =\nu\lim_{t\to\infty} t^{\alpha}\left[(1-\alpha2^{\alpha-1}\kappa\nu) \Bigl(1-\frac{1}{t}\Bigr)^{\alpha}-\kappa\right]\\
 &\geq \nu\left(2^{-\alpha}(1-\alpha2^{\alpha-1}\kappa\nu)-\kappa\right)\lim_{t\to\infty} t^{\alpha}=\infty,
\end{align*}
where in the above steps use was made of $\kappa(2+\alpha \nu)<2^{1-\alpha}$. This implies that $f(t)>0$  for large $t$. As a result, for large $t$, $\nu(t-1)^{\alpha}-1-\kappa\nu t/(t^{1-\alpha}-\alpha\kappa\nu)\geq f(t)/(1-\alpha t^{\alpha-1}\kappa\nu)\geq 0$, from which we know that $T(t,\nu)$  is well-defined for large $t$.
 
   Using the decreasing property of $a(t)$, one can obtain
   \begin{equation}\label{eq:weightineqntemp}
    \int_{t}^{t+T(t,\nu)} s^{-\alpha} ds \leq \sum_{s=t}^{t+T(t,\nu)} a(s)\leq \int_{t-1}^{t+T(t,\nu)} s^{-\alpha} ds.
   \end{equation}
   Consider two cases: (i) $\alpha=1$.  It follows from \eqref{eq:weightineqntemp} that
   \[
   \ln\left(1+\frac{T(t,\nu)}{t}\right) \leq \sum_{s=t}^{t+T(t,\nu)} a(s)\leq \ln\left(1+\frac{T(t,\nu)+1}{t-1}\right).
   \]
    (ii) $0<\alpha<1$.
   In this case, we have
   \[
   \frac{1}{1-\alpha}s^{1-\alpha}\Big|_{t}^{t+T(t,\nu)}\leq \sum_{s=t}^{t+T(t,\nu)} a(s)\leq \frac{1}{1-\alpha}s^{1-\alpha}\Big|_{t-1}^{t+T(t,\nu)}.
   \]
   Employing the mean value theorem, one has $t(t+1)^{-1}\leq \ln(1+t)\leq t$, $\forall t\geq 0$, and for any $\omega>0$, $\omega(1-\alpha)(t+\omega)^{-\alpha}\leq (t+\omega)^{1-\alpha}-t^{1-\alpha}\leq (1-\alpha)\omega t^{-\alpha}$, $\forall t\geq 0$. Combining the previous two cases and the relation $(t+T(t,\nu))^{\alpha}\leq t^{\alpha-1}(t+\alpha T(t,\nu))$ yields
  \begin{equation}\label{eq:weightineqn}
\frac{t^{1-\alpha}T(t,\nu)}{t+\alpha T(t,\nu)}\leq \sum_{s=t}^{t+T(t,\nu)} a(s)\leq    \frac{T(t,\nu)+1}{(t-1)^\alpha},\ \forall t\geq 2,
\end{equation}
  It is easy to verify that \eqref{eq:persistence2} is satisfied with the $\kappa$ and $T(t,\nu)$ given in the proposition.

  Therefore, $a(t)=t^{-\alpha}$ with $0<\alpha\leq 1$ satisfies all the constraints of Theorem~\ref{thm:asyunb}.
  Moreover, it is obvious that condition \eqref{eq:conditionconsis} of Theorem~\ref{thm:consistency} is also satisfied for $0.5<\alpha\leq 1$. This completes the proof.

\section{Proof of Proposition \ref{pro:boundedpower} }\label{app:proofboundedpower}
   Under Assumption~3,  it is clear that  $\mathbb{E}\{\gamma_i(t)\}=0$, $\forall i\in\mathcal{I}_s$ with \eqref{eq:designalphai} and \eqref{eq:designbij}. Hence, $\mathbb{E}\{\mathbf{\Gamma}(t)\}=\mathbf{0}$. It thus follows from Proposition~\ref{pro:propertyL} that $\mathbb{E}\left\{\mathbf{\Gamma}(t)\mathbf{x}(t)\right\}=\mathbb{E}\left\{\mathbf{\Gamma}(t)\right\}\mathbb{E}\left\{\mathbf{x}(t)\right\}=\mathbf{0}$.
  This means that the random channel gain only contributes to the unbiased perturbations $\mathbf{\Gamma}(t) \mathbf{x}(t)$ in \eqref{eq:system},  and $\mathbf{\Gamma}(t) \mathbf{e}(t)$,  $\mathbf{\Gamma}(t) \mathbf{1}$ in \eqref{eq:errdyn}. 

  Now, we turn to the second part of the proposition. By Assumption~3, we know that $\mathbb{E}\{h_{ij}(t)^2\}$ and $\mathbb{E}\{v_{ij}(t)^2\}$ are bounded for all $i,j$. If Assumptions 1, 2 and 4 also hold and $a(t)=t^{-\alpha}$ with $0.5<\alpha\leq 1$, then by  Proposition \ref{pro:weight} and Theorem \ref{thm:consistency}, $\mathbb{E}\{\|\mathbf{x}(t)\|_2^2\}$ is bounded.  

  Considering RN $i\in \mathcal{I}_r$, one can obtain that $R_i(t)\varpropto \sum_{j\in\mathcal{N}_i^s(t)} d_s^{-1} \mathbb{E}\{x_j(t)^2\}$ is  bounded. By the $c_r$ inequality \cite[p.127]{Gut13} and Proposition~\ref{pro:propertyL}, we have  
\begin{align*}
   \mathbb{E}\{u_i(t)^2\}
   &\leq |\mathcal{N}_i^s(t)| \sum_{j\in\mathcal{N}_i^s(t)}\mathbb{E}\left\{(\alpha_j(t) h_{ij}(t)x_j(t)+v_{ij}(t))^{2}\right\}\notag\\
   &=|\mathcal{N}_i^s(t)| \sum_{j\in\mathcal{N}_i^s(t)} \left(d_s^{-1}\mathbb{E}\{h_{ij}(t)^2\}\mathbb{E}\{x_j(t)^2\}
   +\mathbb{E}\{v_{ij}(t)^2\}\right).
\end{align*}
   Hence, the transmit power $P_i(t)=d_r^{-1}\mathbb{E}\{u_i(t)^2\}$ is bounded.

  Next turning to SN $i\in \mathcal{I}_s$, the transmit power $P_i(t)=d_s^{-1}\mathbb{E}\{x_i(t)^2\}$  is bounded and tends to zero as $d_s$ goes to infinity. Furthermore, noting that $\mathbb{E}\{u_j(t)^2\}$ is bounded for all RNs,  SN $i$'s received power $R_i(t)\varpropto \sum_{j\in\mathcal{N}_i^s(t)} d_s^{-1}\mathbb{E}\{x_j(t)^2\}+\sum_{j\in\mathcal{N}_i^r(t)}  d_r^{-1}\mathbb{E}\{u_j(t)^2\}$ is thus bounded.

  Finally, it is easy to see that all the above bounds do not depend on the number of nodes $N$ and the number of SNs $M$.  This completes the proof.

\ifCLASSOPTIONcaptionsoff
  \newpage
\fi


\end{document}